\documentclass[reprint,notitlepage,longbibliography,amsmath,amssymb,onecolumn,prd,aps,superscriptaddress,nofootinbib]{revtex4-1}
\usepackage{graphicx}
\usepackage{microtype}
\usepackage{dcolumn}
\usepackage{float}
\usepackage{acro}
\usepackage{orcidlink}
\usepackage{multirow}
\usepackage{makecell}
\usepackage{booktabs}
\usepackage{subcaption}
\hypersetup{colorlinks=true, allcolors=blue}
\newcommand*{\dif}{\mathop{}\!\mathrm{d}}
\newcommand*{\ee}{\mathrm{e}}
\newcommand*{\ii}{\mathrm{i}}

\DeclareAcronym{GR}{
	short = GR ,
	long = general relativity,
}

\DeclareAcronym{PSD}{
	short = PSD ,
	long = power spectral density,
}

\DeclareAcronym{CBC}{
	short = CBC ,
	long = compact binary coalescence
}

\DeclareAcronym{QNM}{
    short = QNM ,
    long = quasinormal mode
}

\DeclareAcronym{UCO}{
    short = UCO ,
    long = ultracompact object
}

\DeclareAcronym{UniEw}{
    short = UniEw ,
    long = uniform echo waveform model
}

\DeclareAcronym{GW}{
    short = GW ,
    long = gravitational wave
}

\DeclareAcronym{SNR}{
    short = SNR ,
    long = signal-to-noise ratio
}

\DeclareAcronym{ZPK}{
    short = ZPK ,
    long = zero-pole-gain
}

\DeclareAcronym{BH}{
    short = BH ,
    long = black hole
}

\DeclareAcronym{OMC}{
    short = OMC ,
    long = output-mode-cleaner
}

\DeclareAcronym{GMST}{
    short = GMST ,
    long = Greenwich mean sidereal time
}

\DeclareAcronym{ASD}{
    short = ASD ,
    long = amplitude spectral density
}

\DeclareAcronym{LVK}{
    short = LVK ,
    long = LIGO-Virgo-KAGRA collaborations
}

\usepackage[english]{babel}
\usepackage[justification=raggedright]{caption}
\usepackage{comment}

\begin{document}
\title{Model-agnostic search of gravitational wave echoes in LVK data}

\author{Di Wu\orcidlink{0000-0001-7309-574X}}
\email[Contact author: ]{wudi@ucas.ac.cn}
\affiliation{School of Fundamental Physics and Mathematical Sciences, Hangzhou Institute for Advanced Study, UCAS, Hangzhou 310024, People's Republic of China}
\author{Xi-Li Zhang\orcidlink{0009-0003-5329-6247}}
\email{zhangxili@ihep.ac.cn}
\affiliation{Institute of High Energy Physics, Chinese Academy of Sciences, Beijing 100049, People's Republic of China}
\affiliation{School of Physics Sciences, University of Chinese Academy of Sciences, Beijing 100039, People's Republic of China}
\author{Qing-Guo Huang\orcidlink{0000-0003-1584-345X}}
\email{huangqg@itp.ac.cn}
\affiliation{School of Fundamental Physics and Mathematical Sciences, Hangzhou Institute for Advanced Study, UCAS, Hangzhou 310024, People's Republic of China}
\affiliation{School of Physics Sciences, University of Chinese Academy of Sciences, Beijing 100039, People's Republic of China}
\affiliation{Institute of Theoretical Physics, Chinese Academy of Sciences, Beijing 100190, People's Republic of China}
\author{Jing Ren\orcidlink{0000-0002-2116-4659}}
\email[Contact author: ]{renjing@ihep.ac.cn}
\affiliation{Institute of High Energy Physics, Chinese Academy of Sciences, Beijing 100049, People's Republic of China}
\affiliation{Center for High Energy Physics, Peking University, Beijing 100871, People's Republic of China}

\date{\today}
\begin{abstract}
    Gravitational wave echoes offer a unique probe of the near-horizon structure of astrophysical black holes, beyond the standard ``black hole spectroscopy.'' Theoretical waveform predictions, however, remain uncertain, motivating robust searches that avoid specific echo modeling. We present a model-agnostic search framework targeting long-lived quasinormal modes (QNMs) expected from strong interior reflection. By employing a generalized phase-marginalized likelihood that coherently combines data for each QNM across a detector network, our method enhances sensitivity to the signals.
    To handle real detector noise, we implement an optimized notching procedure to suppress instrumental spectral lines and refine the Bayesian parameter settings.
    We validate the performance of this framework using injection studies on O1 background data, demonstrating reliable signal recovery in realistic noise conditions. We then apply this method to three binary black hole merger events with high ringdown signal-to-noise ratios (SNRs): GW150914 from O1, GW231226 from O4a, and the recently reported O4 event GW250114.
    No statistically significant evidence for postmerger echoes is found.
    Consequently, we derive 90\% upper limits on the network SNR and the average initial strain amplitude of the long-lived QNMs.     These results provide model-agnostic constraints on late-time echoes from LVK data, complementing existing searches for other echo signatures.
\end{abstract}

\maketitle

\section{\label{sec:intro}Introduction}

The discovery of \acp{GW} from \ac{CBC} by the \ac{LVK} has opened a new window for exploring the Universe~\cite{LIGOScientific:2016aoc}. To date, more than $200$ \ac{CBC} events have been detected~\cite{LIGOScientific:2018mvr,LIGOScientific:2020ibl,KAGRA:2021vkt,LIGOScientific:2025slb}.
Among these, the recent detection of GW250114 during the \ac{LVK} O4 observing run stands as a notable milestone: its high ringdown \ac{SNR} has enabled precise black hole spectroscopy~\cite{LIGOScientific:2025rid,LIGOScientific:2025wao}.
While the early ringdown waveforms are broadly consistent with the predictions of \ac{GR} for Kerr \acp{BH}~\cite{Isi:2019aib,Capano:2021etf,Cotesta:2022pci,Wang:2024jlz,Giesler:2024hcr,Berti:2025hly}, they remain largely insensitive to near-horizon corrections, as the large gravitational redshift near the horizon would push potential deviations to much later times. If the merger remnant is an \ac{UCO} with a reflective surface located just outside the would-be horizon, as suggested by various quantum-gravity inspired models, \acp{GW} would be trapped between the light-ring potential barrier and this interior boundary. The repeated, gradual leakage of waves through the barrier produces a series of quasiperiodic wavelets following the initial ringdown signal, known as gravitational wave echoes~\cite{Cardoso:2016rao,Cardoso:2016oxy,Bueno:2017hyj,Zhang:2017jze,Urbano:2018nrs,Mannarelli:2018pjb,Li:2019kwa,Cardoso:2019apo,Conklin:2019fcs,Wang:2019rcf,Holdom:2020onl,Fang:2021iyf,Vlachos:2021weq,Zhang:2021fla,Huang:2021qwe,Ikeda:2021uvc,Konoplya:2022zym,Abedi:2022omf,Yang:2022ryf,Vellucci:2022hpl,Guo:2022umh,Mukherjee:2022wws,Oshita:2023tlm,Saketh:2024ojw,Yang:2024rms,Yang:2024prm,Zhu:2024gvl,Lin:2025lsz,Cao:2025qxt,Hu:2025beh,Terashima:2025tct}.
Notably, the time delay for the echo wavelets remains observationally accessible even for the extreme case in which the reflective surface lies only a Planck length outside the would-be horizon.

However, searching for these echoes presents significant challenges due to theoretical uncertainties. Various model-dependent templates were developed to capture the echo waveforms, and some have been applied to real-data searches~\cite{Abedi:2016hgu,Abedi:2017isz,Westerweck:2017hus,Maselli:2017tfq,Lo:2018sep,Nielsen:2018lkf,Uchikata:2019frs,Wang:2019szm,Wang:2020ayy,Abedi:2021tti,Westerweck:2021nue,Abedi:2022bph,Uchikata:2023zcu,Lai:2025skp,Mondal:2025uxm,Hamoudy:2025cjm,Lai:2026yvm}, but they often rely on specific assumptions regarding the \ac{UCO} structure that are not yet well understood. To overcome this limitation, model-agnostic search methods have been proposed~\cite{Conklin:2017lwb,Tsang:2018uie,Abedi:2018npz,Tsang:2019zra,Salemi:2019uea,Holdom:2019bdv,Conklin:2021cbc,Ren:2021xbe,Miani:2023mgl,Wu:2023wfv}.
A time-domain approach employs pipelines developed for generic burst signals such as \texttt{coherent WaveBurst}~\cite{Salemi:2019uea,Conklin:2021cbc,Miani:2023mgl} or \texttt{BayesWave}~\cite{Tsang:2018uie,Tsang:2019zra} to search for the first few wavelets following the main event. This strategy is primarily sensitive to short-lived \acp{QNM} of the \ac{UCO},
which occur either when the interior reflection is weak or above the \ac{BH} ringdown frequency, where the light-ring potential barrier reflects inefficiently.
Conversely, strong interior reflections cause the low-frequency content of the echo to decay slowly, allowing the \ac{SNR} to accumulate steadily over time. Capturing such extended signals requires analyzing many wavelets, which renders burst-based searches computationally inefficient. In this regime, a frequency-domain approach is more appropriate, as it directly targets the long-lived \acp{QNM} that dominate late-time echoes and appear as narrow, quasiperiodic resonances across a wide frequency band.

A uniform comb model with narrow line widths can capture the generic features of these long-lived \acp{QNM}~\cite{Conklin:2017lwb,Holdom:2019bdv}.
By marginalizing over the unknown phase information, an efficient likelihood function was derived by some of the present authors to search for these modes in a model-agnostic framework. This method was applied to O1 data, and it found no clear evidence of narrow line structures for the first two CBC events~\cite{Ren:2021xbe}. However, this earlier approach completely discarded phase information, limiting its sensitivity to weak signals.
More recently, a new phase-marginalized likelihood retaining partial phase information was proposed to address this sensitivity limitation~\cite{Wu:2023wfv}. By accounting for the relative phase variation between different frequency bins belonging to the same \ac{QNM}, this new method was shown to yield significant performance improvements in single-detector Gaussian noise. In particular, it delivers robust search results when the analysis duration is varied, especially in the high-frequency resolution regime.

In this work, we extend the new phase-marginalized likelihood to a network of multiple detectors, enabling the coherent combination of data across the network to enhance detection sensitivity. We also improve the real-data search pipeline in several key aspects relative to earlier studies~\cite{Ren:2021xbe,Wu:2023wfv}. These improvements include an optimized notching procedure to suppress the dominant non-Gaussian artifacts---instrumental spectral lines---as well as more appropriate settings for intrinsic and extrinsic parameters. We validate the performance of this pipeline on O1 detector noise by comparing with Gaussian noise results, confirming the effectiveness of notching and the robustness of signal recovery.
We then apply this robust pipeline to search for gravitational wave echoes in \ac{LVK} events, focusing on three events that exhibit high ringdown \ac{SNR}: GW150914 from O1, GW231226\_101520 (hereafter GW231226) from O4a, and the recently reported O4 event GW250114\_082203 (hereafter GW250114).

This paper is structured as follows: in Sec.~\ref{sec:likelihood}, we review the \acp{QNM} of \acp{UCO} in the case of strong interior reflection and derive the phase-marginalized likelihood for a multidetector network, incorporating the individual detector responses. The data analysis procedure and parameter settings for \ac{LVK} events are detailed in Sec.~\ref{sec:lvk}.
Section ~\ref{sec:inject} presents injection tests that validate the performance of this pipeline on real detector noise.
In Sec.~\ref{sec:result}, we apply our search pipeline to the three selected \ac{LVK} events and present the search results. We summarize in Sec.~\ref{sec:summary}. Additional information on instrumental lines relevant to the search and supplementary results for pure Gaussian noise are provided in Appendixes~\ref{app:line} and \ref{app:gaussian}.
Throughout the paper, we adopt geometric units, $G=c=\hbar=1$, unless stated otherwise.

\section{Phase-marginalized likelihood for a multidetector network}\label{sec:likelihood}
\subsection{General features of the echo waveform}\label{subsec:waveform}

Candidates for \acp{UCO} have been proposed both in \ac{GR} with exotic matter and in modified theories of gravity (see, e.g., the review in Ref~\cite{Cardoso:2019rvt}). Some constructions show explicit connection between \acp{UCO} and quantum gravity effects~\cite{Mathur:2005zp,Holdom:2016nek, Ren:2019afg,Holdom:2019ouz}, and the resulting echo signatures have been studied in various models~\cite{Holdom:2020onl,Ikeda:2021uvc}. For a model-agnostic discussion, we adopt a simple phenomenological model: a truncated Kerr \ac{BH} with a reflective surface located just outside the would-be horizon~\cite{Conklin:2017lwb,Nakano:2017fvh,Wang:2019rcf,Maggio:2019zyv, Xin:2021zir,  Srivastava:2021uku,Vellucci:2022hpl,Deppe:2024fdo,Rosato:2025byu, Rosato:2025lxb}. Although simplified, this model captures a wide range of \ac{UCO} predictions by varying the surface location and its effective reflectivity, thereby providing an efficient phenomenological framework for the subsequent analysis. In this setup,
the average time delay $t_d$ between adjacent echo wavelets is approximately twice the light travel time between the light-ring potential barrier and the reflective surface~\cite{Cardoso:2019rvt,Berti:2025hly}
\begin{equation}\label{eq:td}
    \frac{t_d}{M}\approx 2\left[1+(1-\chi^2)^{-1/2}\right] \ln\left(\frac{1}{\epsilon}\right)\,,
\end{equation}
where $M$ and $\chi$ are the mass and dimensionless spin of the \ac{UCO}, respectively. The parameter $\epsilon$ characterizes the location of the reflective surface $r_0$ via $r_0=r_+(1+\epsilon)$,
where $r_+$ denotes the \ac{BH} horizon radius.

In the frequency domain, for a given angular harmonic $(\ell, m)$, the echo waveform for this simplified \ac{UCO} model can be written as~\cite{Xin:2021zir,Wu:2023wfv}:
\begin{equation}\label{eq:echoeff}
    h_{\mathrm{echo}}(\omega) = \mathcal{P}(\omega) h_{\mathrm{eff}}(\omega) \, , \quad \mathcal{P}(\omega) = \frac{\mathcal{R}_{\mathrm{BH}}(\omega) \mathcal{R}_{\mathrm{wall}}(\omega)}{1 - \mathcal{R}_{\mathrm{BH}}(\omega) \mathcal{R}_{\mathrm{wall}}(\omega)} \, ,
\end{equation}
where $\mathcal{R}_{\mathrm{wall}}(\omega)$ represents the effective reflectivity of the \ac{UCO} inner boundary at $r_0$, and $\mathcal{R}_{\mathrm{BH}}(\omega)$ is the reflectivity of the \ac{BH} potential barrier for the given $(\ell, m)$, calculated following Ref.~\cite{Oshita:2022pyf}. The quantity $h_{\mathrm{eff}}(\omega)$ denotes the \ac{BH} response at the horizon and carries the source information.
The processing function $\mathcal{P}(\omega)$ depends on the combined reflectivity of the cavity, i.e. $\mathcal{R}_{\rm BH}(\omega)\mathcal{R}_{\rm wall}(\omega)\equiv \mathcal{R}_{\mathrm{eff}}(\omega)e^{i\delta(\omega)}$, where $\mathcal{R}_{\mathrm{eff}}(\omega)$ and $\delta(\omega)$ are its amplitude and phase, respectively. Because $|\mathcal{R}_{\rm BH}(\omega)|\ll1$ above the light-ring barrier, the amplitude $\mathcal{R}_{\mathrm{eff}}(\omega)$ is unsuppressed only for frequencies below the \ac{BH} ringdown frequency $f_{\rm RD}$.\footnote{We require $|\mathcal{R}_{\mathrm{wall}}|$ to be sufficiently below unity so that $\mathcal{R}_{\mathrm{eff}}\lesssim 1$, thereby avoiding the ergoregion instability (see, e.g., Refs.~\cite{Friedman:1978ygc, Cardoso:2007az, Maggio:2017ivp, Maggio:2018ivz} for detailed discussions).} The phase $\delta(\omega)$ encodes the propagation-time dependence. For a \ac{UCO} with $t_d$ much longer than the timescale $\sim M$, we can approximate $\delta(\omega)\approx t_d(\omega-\omega_H)$, where $\omega_H=m\chi/(2r_+)$. The poles of $\mathcal{P}(\omega)$ in the complex-frequency plane give the \acp{QNM} of \acp{UCO}, which are labeled by the angular harmonic $(\ell, m)$ and a radial number $n$ that enumerates the radial eigensolutions~\cite{Mark:2017dnq,Macedo:2018yoi,Maggio:2019zyv,LongoMicchi:2019wsh,Conklin:2019fcs}. When interior reflection is strong, the combined reflectivity $\mathcal{R}_{\mathrm{eff}}(\omega)$ can approach unity for $f<f_{\rm RD}$. Following the derivation presented in Ref.~\cite{Wu:2023wfv} (see in particular Eq.~(5) therein),  the \acp{QNM} in this low-frequency regime can be obtained analytically from  $1-\mathcal{R}_{\mathrm{eff}}(\omega)e^{i\delta(\omega)}=0$, which yields a series of characteristic modes $(\omega_R,\omega_I)=(2\pi f_n, -1/\tau_n)$ satisfying
\begin{equation}\label{eq:fntaun}
    t_d(f_{n+1}-f_n)\approx 1,\quad
    t_d/\tau_n\approx
    -\ln \mathcal{R}_{\mathrm{eff}}(f_n)\,.
\end{equation}
These are essentially the quasiperiodic trapped modes of the cavity, with a roughly constant spacing set by the echo time delay, $\Delta f \sim 1/t_d$ to leading order, and a lifetime much longer than $t_d$ depending on how closely  $\mathcal{R}_{\rm eff}$ approaches unity. They dominate the late-time behavior of postmerger echoes and can accumulate a large \ac{SNR} over a sufficiently long observation. For these long-lived, quasiperiodic \acp{QNM},
the echo waveform near each modal frequency is dominated by a single-mode contribution.
Considering the realistic analysis of a finite data segment,
the amplitude and phase around the $n$-th mode can be approximated by \cite{Wu:2023wfv}
\begin{eqnarray}
	\left|{h_{\rm echo}^{(T)}(f)} \right| & \approx & \frac{A_n}{\sqrt{4\pi^2(f-f_n)^2+1/\tau^2_n}} \left|{1-\ee^{-T/\tau_n}\ee^{\ii 2\pi (f-f_n)T}} \right| \, , \label{eq:abshf}\\
	\arg \left(h_{\rm echo}^{(T)}(f)  \right)  & \approx & \arg \left(\frac{1-\ee^{-T/\tau_n}\ee^{\ii 2\pi (f-f_n)T}}{\ii 2\pi(f-f_n)+ 1/\tau_n}\right) +  \left(\delta_n+ 2\pi f t_d\right) \, . \label{eq:arghf}
\end{eqnarray}
Here, $A_n$ is the time-domain strain amplitude of the $n$-th mode  at the start of the analyzed data segment, or at the onset of the mode if it is excited later;
$\delta_n$ its constant phase, and $T$ is the finite duration of the strain data.
The first term in the phase originates from the Lorentzian line shape, which produces a rapid phase shift of approximately $\pi$ across the narrow resonance width of $\sim 1/(\pi\tau_n)$. In comparison, the second term $\delta_n+2\pi f t_d$ changes weakly across the same interval, by an amount $\sim 2\pi t_d/(\pi\tau_n) \sim 2 t_d/\tau_n$. Using Eq.~(\ref{eq:fntaun}), one finds $t_d/\tau_n \approx -\ln \mathcal{R}_{\rm eff}(f_n) \ll 1$ in the strong-reflection regime, so this variation is negligible compared to the $\pi$ phase shift.

\begin{figure}[htbp]
	\centering
	\includegraphics[width=12cm]{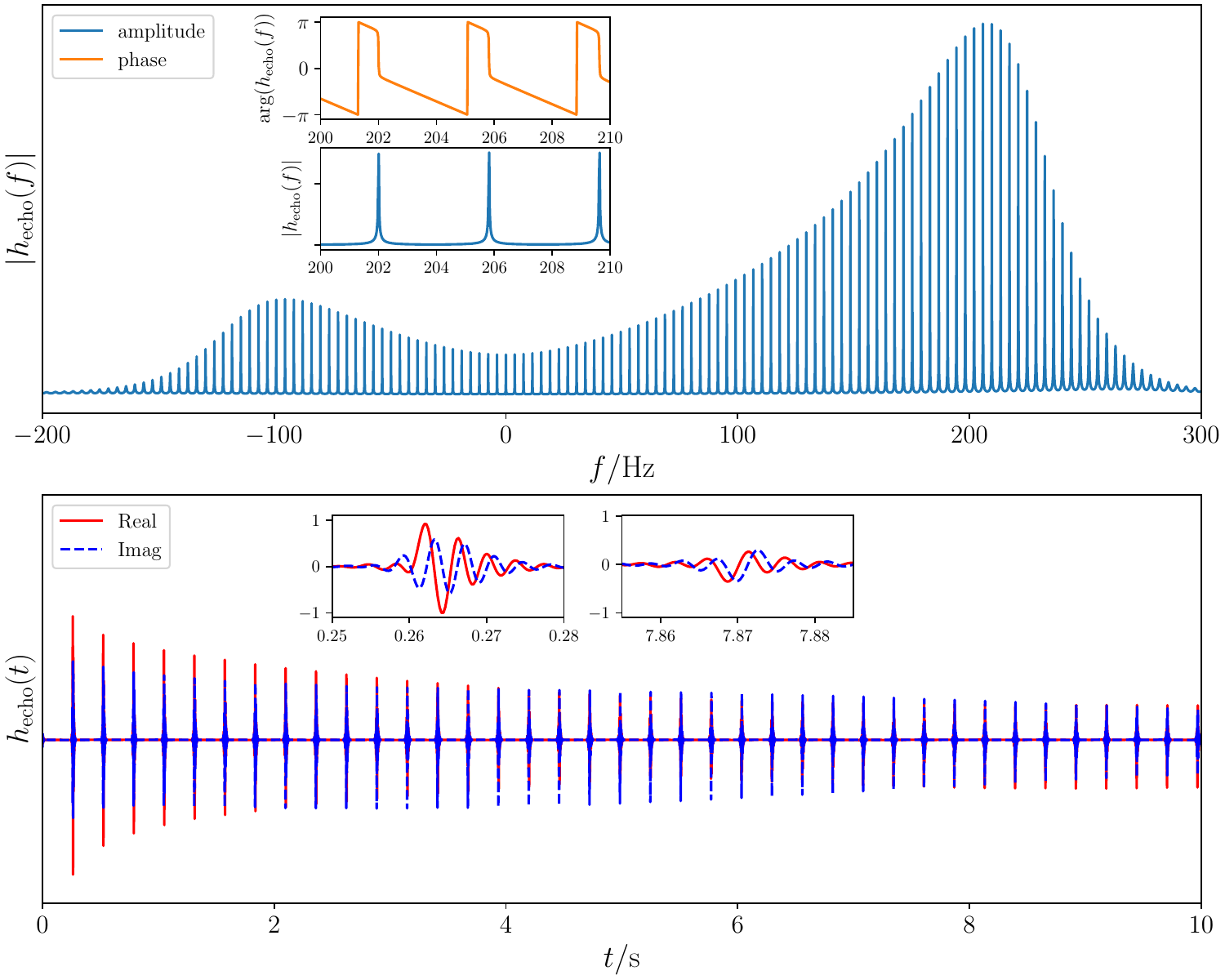}
	\caption{\label{fig:waveform}{A representative benchmark echo waveform for the truncated Kerr \ac{BH} model in the frequency domain (top panel) and time domain (bottom panel), with $M=62\,M_{\odot}$, $\chi=0.67$, $t_d\approx (1/3.81) $\,s, $T\approx49$\,s, $\mathcal{R}_{\mathrm{wall}}(\omega)=0.99$ and $h_{\mathrm{eff}}(\omega)=h_{\mathrm{RD}}(\omega)$, where $h_{\mathrm{RD}}$ represents the ringdown excitation of the dominant $l = m = 2$ fundamental mode \cite{Maggio:2019zyv} and $M_{\odot}$ is the solar mass. The top panel inset shows a zoomed-in view around the peak of amplitude. The blue and orange curves represent the absolute value and the phase of the waveform, respectively. The bottom panel displays a zoomed-in view around the $1$-st and the $30$-th echo wavelets. The red solid and blue dashed curves represent the real part and the imaginary part of the waveform, respectively.}}
\end{figure}

For illustration, Figure~\ref{fig:waveform} displays the frequency-domain and time-domain waveforms of a representative benchmark echo signal in the strong-reflection case. The narrow quasiperiodic resonances originate from the pole structure of the processing function $\mathcal{P}(\omega)$ in Eq.~(\ref{eq:echoeff}).
The inset in the top panel shows the frequency-domain amplitude and phase variations around a few long-lived \acp{QNM}. Each narrow Lorentzian resonance produces a peak in the amplitude together with a sharp, nearly $\pi$-sized phase shift, corresponding to the first term in Eq.~(\ref{eq:arghf}). In addition, a slow, nearly linear trend arises from the second term in Eq.~(\ref{eq:arghf}), which is negligible within the narrow width of the \acp{QNM}. The overall envelope of the frequency domain amplitude is further modulated by the frequency content of $h_{\rm eff}(\omega)$.

The properties of the \acp{QNM} depend crucially on both the effective reflection at the interior boundary and the frequency content of the initial pulse, as given in  Eq.~(\ref{eq:echoeff}). For \acp{UCO} with $1/t_d\ll f_{\rm RD}$, a large number of long-lived \acp{QNM} can typically be excited. Searching for all modes individually then becomes challenging because their spacing, width, and amplitude can vary from mode to mode.
Nonetheless, because in the strong-reflection limit the \acp{QNM} form a quasiperiodic trapped-mode spectrum whose leading-order spacing is set by the echo time delay $\Delta f \approx 1/t_d$ from Eq.~(\ref{eq:fntaun}),
we showed in Ref.~\cite{Wu:2023wfv} that a simple \ac{UniEw} can be used for a model-agnostic search. By treating the spacing, amplitude, and width of the modes as constants, this template can efficiently capture the dominant \acp{QNM} for a wide range of theoretical predictions, and its parameters provide a useful estimate of these average mode properties. Because we will later marginalize over the approximately constant phase term $\delta'_n$ for each mode, we construct the \ac{UniEw} template from Eqs.~(\ref{eq:abshf}) and (\ref{eq:arghf}) by keeping only the first phase term,
\begin{equation}\label{eq:uniformQNM}
    \tilde{h}(f)= \sum_{n=N_{\rm min}}^{N_{\rm max}} \left\{\begin{aligned}
		&A^\prime \frac{-\ii/\tau}{2\pi(f-f_n)-\ii/\tau}\left(1-\ee^{-T/\tau}\ee^{\ii 2\pi (f-f_n)T}\right), &|f-f_n|\leq f_\textrm{cut} ,\\
		&0, &|f-f_n|> f_\textrm{cut} .
	\end{aligned}
	\right.
\end{equation}
This template is defined in the frequency band $[f_{\rm min}, f_{\rm max}]$. The central frequencies of the equally spaced \acp{QNM} are given by $f_n=\Delta f(n+q_0)$, where the index runs from $n=N_{\rm min}$ to $N_{\rm max}$, determined by $N_{\rm min}=\lceil f_{\rm min}/\Delta f \rceil$ and $N_{\rm max}=\lfloor f_{\rm max}/\Delta f \rfloor$. The coefficient $A^\prime=A \tau$ represents the mode-averaged peak amplitude in the frequency domain, with $A$ being the average of the initial time-domain strain amplitudes $A_n$ in Eq.~(\ref{eq:abshf}).\footnote{Ref.~\cite{Wu:2023wfv} defines the \ac{UniEw} model with the time-domain amplitude $A$, but the Bayesian search parameter is  the frequency-domain amplitude $A^\prime$, with the same unit as $h(f)$. The notation does not explicitly separate $A^\prime$ from $A$. But this typo does not affect the analysis or the reported results.}
A cutoff $f_\textrm{cut}$ is imposed to ensure that each mode is distinct and well separated.

\subsection{Phase-marginalized likelihood construction}\label{subsec:likelihood}
As proposed in Ref.~\cite{Wu:2023wfv}, an efficient search strategy for these \acp{QNM} is to retain only the first phase term in Eq.~(\ref{eq:arghf}), which dominates the phase variation and is determined by the same parameter set as the amplitude in Eq.~(\ref{eq:abshf}), while treating the slowly varying second term as a constant. This constant phase is then marginalized in the likelihood function as a nuisance parameter. In Ref.~\cite{Wu:2023wfv}, we derived the phase-marginalized likelihood for a single GW detector without including the detector response. Here, we generalize this derivation to a network of multiple detectors, incorporating their respective detector responses.
In the frequency domain, the strain data for the $I$-th detector can be modeled as: $d_I(f)=n_I(f)+R_I(f)h(f)$, where $n_I$ is stationary Gaussian noise, $h(f)$ is the signal waveform, and $R_I(f)$ denotes the effective impulse response for the $I$-th detector.\footnote{The spinning-UCO waveform $h(f)$ is asymmetric in frequency. Including the detector response gives the real time-domain strain $h_I(f)=\frac{1}{2}[R_I(f)h(f)+R^*_I(f)h^*(-f)]$, which projects both positive and negative frequency parts onto the positive side, resulting in a two-component structure~\cite{Conklin:2019fcs,Ren:2021xbe}. Since our search selects only the dominant component, we use the simplified notation $h_I(f)=R_I(f)h(f)$ here.}
For a given frequency bin, the Gaussian likelihood in polar coordinates is given by
\begin{equation}\label{eq:GaussianL1}
   \mathcal{L}_I(d_{I,j}|h_{j}) =\frac{|d_{I,j}|}{2\pi P_{I,j}}\exp\left(-\frac{1}{2}\frac{|d_{I,j}|^2+|R_{I,j}h_{j}|^2}{\tilde{P}_{I,j}}\right) \exp\left(\frac{|d_{I,j}||R_{I,j} h_{j}|}{\tilde{P}_{I,j}}\cos(\phi_{I,j}-\theta_{I,j}-\delta_{j})\right)\,,
\end{equation}
where, $d_{I,j}$, $R_{I,j}$, and $h_{j}$ denote the data, impulse response, and signal model in the $j$-th frequency bin; $\phi_{I,j}$, $\theta_{I,j}$, and $\delta_{j}$ represent their corresponding phases. $P_{I,j}$ is the one-sided \ac{PSD}, and $\tilde{P}_{I,j} = P_{I,j}/4\delta f$ is the normalized one-sided \ac{PSD}, with $\delta f=1/T$ being the frequency resolution. Since the noise is independent across frequency bins and detectors, the joint likelihood is the product of individual likelihoods.
We divide all frequency bins within the band into distinct subsets $\{\mathcal{J}_n\}$, where each subset $\mathcal{J}_n$ contains the bins around the $n$-th \ac{QNM}. We then combine the likelihoods for all bins $j \in \mathcal{J}_n$ associated with the $n$-th mode and marginalize over the approximately constant phase term $\delta^\prime_n\equiv \delta_n+2\pi f t_d$.
The marginalized likelihood is then expressed as
\begin{equation}
	\begin{split}\label{eq:magrlikelihood}
    \mathcal{L}_{n}(d|h) = &\int_0^{2\pi} \left(\prod_{I,j\in \mathcal{J}_n}\mathcal{L}_I(d_{I,j}|h_j)\right)\pi(\delta^\prime_n)\dif \delta^\prime_n \\
    = &\left(\prod_{I, j\in \mathcal{J}_n}\frac{|d_{I,j}|}{2\pi P_{I,j}}\right)\exp\left(\sum_{I, j\in \mathcal{J}_n}-\frac{1}{2}\frac{|d_{I,j}|^2+|R_{I,j}h_{j}|^2}{\tilde{P}_{I,j}}\right) \\
    &\int^{2\pi}_0\exp\left(\sum_{I, j\in \mathcal{J}_n}\frac{|d_{I,j}||R_{I,j}h_{j}|}{\tilde{P}_{I,j}}\cos(\varphi_{I,j}-\delta^\prime_n)\right)\pi(\delta^\prime_n)\dif \delta^\prime_n \\
    = &\left(\prod_{I, j\in \mathcal{J}_n}\frac{|d_{I,j}|}{2\pi P_{I,j}}\right)\exp\left(\sum_{I, j\in \mathcal{J}_n}-\frac{1}{2}\frac{|d_{I,j}|^2+|R_{I,j}h_{j}|^2}{\tilde{P}_{I,j}}\right)I_0\left(\left|\sum_{I, j\in \mathcal{J}_n}\frac{|d_{I,j}||R_{I,j}h_{j}|}{\tilde{P}_{I,j}}\ee ^{\ii\varphi_{I,j}}\right|\right) \\
    = &\left(\prod_{I, j\in \mathcal{J}_n}\frac{|d_{I,j}|}{2\pi P_{I,j}}\right)\exp\left(\sum_{I, j\in \mathcal{J}_n}-\frac{1}{2}\frac{|d_{I,j}|^2+|R_{I,j}h_{j}|^2}{\tilde{P}_{I,j}}\right)I_0\left(\left|\sum_{I, j\in \mathcal{J}_n}\frac{d_{I,j}(R_{I,j}h_{j})^*}{\tilde{P}_{I,j}}\right|\right)\,,
	\end{split}
\end{equation}
where $\pi(\delta^\prime_n) = 1/(2\pi)$ is the uniform prior for the approximately constant phase, and $\varphi_{I,j}=\phi_{I,j}-\theta_{I,j}-(\delta_j-\delta^\prime_n)$ denotes the residual phase offset in detector $I$ and frequency bin $j$ after separating out $\delta^\prime_n$.
The integral over $\delta^\prime_n$ is evaluated using the identity $\int^{2\pi}_0\exp\left[\sum_i a_i\cos(x+b_i)\right]\dif x=2\pi I_0\left(\left|\sum_i a_i\ee^{\ii b_i}\right|\right)$,
which analytically marginalizes over the constant phase $\delta^\prime_n$ and yields the zeroth-order modified Bessel function of the first kind, $I_0$~\cite{Veitch:T1300326v1,Veitch:2014wba,Thrane:2018qnx}.
The last line rewrites the Bessel-function argument compactly in terms of the full waveform $h_j$:
the argument differs from the preceding expression only by the common phase factor $\ee^{\ii\delta'_n}$ for all $j\in\mathcal{J}_n$, whose modulus is unity and therefore drops out inside the absolute value.
The final expression shows that all frequency bins belonging to the same mode,
as well as data from different detectors, enter coherently through the complex sum inside the argument of $I_0$.

Combining all \acp{QNM}, the phase-marginalized log-likelihood normalized by the noise contribution is given by:
\begin{equation}
	\begin{split}
		\ln \mathcal{L}&\equiv \sum_{n} \left[\ln \mathcal{L}_n(d| h)- \ln \mathcal{L}_n(d|0)\right] \\
        &= {\left[{\sum_{n} \ln I_0\left(\left|\sum_{j\in \mathcal{J}_n} \left(h_{j}^*\sum_I\frac{d_{I,j}R_{I,j}^*}{\tilde{P}_{I,j}}\right)\right|\right)}\right]} -\frac{1}{2} \sum_{I} \sum_{j} \frac{|R_{I,j}h_{j}|^2}{\tilde{P}_{I,j}}\, .
	\end{split}
\end{equation}
Here, the first term measures the signal-data overlap with partial phase information, while the second gives the network optimal \ac{SNR}.
For further simplification, it is useful to define the combined data $\tilde{d}_j$ and the effective search template $\tilde{h}_j$ as
\begin{eqnarray}\label{eq:combdata}
    \tilde{d}_j = \sum_Id_{I,j}  \frac{\tilde{P}_{j}}{\tilde{P}_{I,j}}\frac{R^*_{I,j}}{R^*_{J,j}},\quad
    \tilde{h}_j=h_jR_{J,j}\,,
\end{eqnarray}
where $J$ denotes a reference detector with non-negligible response and $\tilde{P}_j^{-1}=\sum_I (|R_{I,j}|^2/|R_{J,j}|^2)\tilde{P}^{-1}_{I,j}$ represents the combined normalized \ac{PSD}. The log-likelihood can then be written as:
\begin{equation}\label{eq:newlnL}
    \ln\mathcal{L} = {\left[{\sum_{n}\ln I_0\left(\left|\sum_{j\in \mathcal{J}_n}\frac{\tilde{d}_j\tilde{h}^*_j}{\tilde{P}_j}\right|\right)}\right]}-\frac{1}{2}\sum_{j}\frac{|\tilde{h}_j|^2}{\tilde{P}_j}\,.
\end{equation}
We refer to this expression as the \textit{new likelihood} in the following discussion. Notably, it matches precisely the single-detector likelihood derived earlier in Ref.~\cite{Wu:2023wfv}. Since the combined \ac{PSD} for the detector network is lower than that of the reference detector alone, the coherent combination is expected to enhance search sensitivity. When an injected echo signal is present and the search template matches the injection, the overlap term $\sum_{j\in \mathcal{J}_n}\tilde{d}_j\tilde{h}^*_j/\tilde{P}_j$ is dominated by $\sum_{j\in \mathcal{J}_n}|\tilde{h}_j|^2/\tilde{P}_j$, considering that noise contributions largely average out due to random phase cancellation across frequency bins. Therefore, much like in the single-detector case, the phase-marginalized likelihood for a given mode is sensitive to its network optimal \ac{SNR}---a result of coherently combining all frequency bins associated with that mode. In contrast, the full likelihood combines different \acp{QNM} noncoherently.

For comparison, we also present the earlier form of the phase-marginalized likelihood derived in Ref.~\cite{Ren:2021xbe}, in which the phase was marginalized independently for each frequency bin. The corresponding normalized log-likelihood for a detector network is given by~\cite{Ren:2021xbe}
\begin{equation}\label{eq:oldlnL}
    \ln\mathcal{L}_{\rm old} = \sum_{j}\ln I_0\left(\left|\frac{\tilde{d}_j\tilde{h}^*_j}{\tilde{P}_j}\right|\right)-\frac{1}{2}\sum_{j}\frac{|\tilde{h}_j|^2}{\tilde{P}_j}\,.
\end{equation}
We refer to this as the \textit{old likelihood} in later discussions.
The conceptual difference between Eqs.~(\ref{eq:newlnL}) and~(\ref{eq:oldlnL}) lies in how the unknown phase is treated. In Eq.~(\ref{eq:newlnL}), the slowly varying phase term $\delta^\prime_n$ is assumed to be approximately constant within each \ac{QNM}, so the phase is marginalized once per mode and all bins $j\in\mathcal{J}_n$ are combined coherently through the complex sum inside $I_0$. In Eq.~(\ref{eq:oldlnL}), by contrast, the phase is marginalized independently in each frequency bin. This removes the coherent combination across neighboring bins and makes the statistic depend mainly on the \ac{SNR} per bin.
In the low-resolution limit, the \ac{SNR} for a single \ac{QNM} is dominated by just one frequency bin, so the coherent sum in Eq.~(\ref{eq:newlnL}) receives a dominant contribution from only that term. The relative-phase information across bins therefore becomes irrelevant, and the result reduces to the form in Eq.~(\ref{eq:oldlnL}).

For the postmerger search, the response function can in principle be inferred from inspiral-merger-ringdown analyses of the main event, though its precise form remains weakly constrained due to strong degeneracies among extrinsic parameters. Nonetheless, these uncertainties have limited impact on our search for long-lived \acp{QNM} of \acp{UCO}.
For ground-based interferometers in the small-antenna limit, the response function can be well approximated as~\cite{Romano:2016dpx}
\begin{equation}
	R^{ab}_I(f,\hat{n}) \approx \frac{1}{2}\left(u^a_Iu^b_I-v^a_I v^b_I\right) \ee^{\ii \,2\pi f\, \hat{n}\cdot \vec{x}_I/c} \, ,\label{rf}
\end{equation}
where $\hat{n}$ is the unit vector pointing to the gravitational wave source, $u^a_I$ and $v^a_I$ are the unit vectors along the two arms of detector $I$ and $x_I$ is the location of detector vertex \cite{ligo:1998coord}.
Since frequency dependence enters only through the phase term---which varies slowly near the central frequency of the \acp{QNM} compared to the first term in Eq.~(\ref{eq:arghf})---the effective search template $\tilde{h}_j$ incorporating the response [see Eq.~(\ref{eq:combdata})] differs only mildly from the original signal waveform $h_j$. To enable an efficient, model-agnostic search for \acp{QNM} from various theoretical scenarios, we adopt a simplified search template for these modes~\cite{Wu:2023wfv}, as detailed below. This simplified model can then be used here as the effective search template.
The combined data in Eq.~(\ref{eq:combdata}) can be written as
\begin{equation}\label{eq:dIJ}
    \tilde{d}_j = \tilde{P}_{j} \left( \frac{d_{J,j}}{\tilde{P}_{J,j}}+ \sum_{I\neq J} A_{JI} \ee^{\ii \phi_{JI,j}}\frac{d_{I,j}}{\tilde{P}_{I,j}} \right) \,,
\end{equation}
where $R_{I,j}/R_{J,j}\equiv A_{JI} \ee^{-\ii \phi_{JI,j}}$, with $A_{JI}$ and $\phi_{JI,j}$ representing the relative amplitude and phase of the responses between the two detectors.
The relative phase can be further expressed as
\begin{equation}\label{eq:phiHL}
	\phi_{JI,j}=\phi_{JI,0}-2\pi f_j \Delta t_{JI} \,,
\end{equation}
where $\phi_{JI,0}$ is a constant phase term and $\Delta t_{JI}=t_J-t_I$ is the lag of signal arrival time between each pair of detectors.
The detector-response  parameters $A_{JI}$, $\phi_{JI,0}$, and $\Delta t_{JI}$ all depend on the source sky location. In a general network with many detectors, the most consistent Bayesian treatment is to express them as functions of the sky position and to set their priors using the sky-location posterior from the main event analysis. However, because $\Delta t_{JI}$ is typically well determined in the main event detection, a practical minimal approach for the present study---where most events have significant \ac{SNR} in fewer than three detectors---is to fix $\Delta t_{JI}$ as the best-fit value and  treat $\{A_{JI}, \phi_{JI,0}\}$ as free parameters with appropriate priors, depending on the specific detector network configuration.

\section{LVK data analysis pipeline}\label{sec:lvk}

Observational detector data are neither strictly stationary nor Gaussian. Our study focuses on  signals with durations of order $\mathcal{O}(10)-\mathcal{O}(100)$\,s; over such intervals, the strain data can be considered approximately stationary. The dominant non-Gaussian effects typically originate from narrow spectral lines caused by instrumental or environmental disturbances. These lines can mimic the characteristic long-lived \acp{QNM} associated with echoes, owing to the similarity between the Lorentzian line profile in Eq.~(\ref{eq:abshf}) and typical instrumental line shapes~\cite{Ren:2021xbe,LSC:2018vzm,LIGO:2024kkz,LIGOScientific:2025snk}.
In Ref.~\cite{Ren:2021xbe}, using the old likelihood function in Eq.~(\ref{eq:oldlnL})---which discards all phase information---we found that instrumental lines in O1 data produced a long tail of noise outliers in the background distribution. Nevertheless, most of these outliers can be efficiently removed by notching out a small number of the strongest instrumental lines in the catalogs.
With the new likelihood in Eq.~(\ref{eq:newlnL}), which retains relative phase information, we expect the search to be less susceptible to contamination by instrumental lines. The reason is that the phase variation of most instrumental lines is likely to differ considerably from the signal model in Eq.~(\ref{eq:arghf}); therefore, coherent combination of nearby frequency bins would lead to substantial cancellation, provided the frequency resolution is sufficiently high.
However, as discussed below Eq.~(\ref{eq:oldlnL}), the new likelihood reduces exactly to the old one when a resonance peak contains only one frequency bin, and Eq.~(\ref{eq:oldlnL}) remains a good approximation when a resonance is unresolved or only weakly resolved. Moreover, some lines may still be captured due to coincidental phase similarity or random noise fluctuations. We therefore continue to apply notch filtering in the present analysis.
It is worth noting that later observing runs contain significantly fewer instrumental lines than O1. As a result, in the frequency band relevant to the present search, the number of dominant lines that actually need to be notched is typically only of order a few.

Concretely, for a given time duration $T$, we implement the following steps to prepare the frequency-domain data $d_{I,j}$ for the likelihood function.
First, we select time-series points that satisfy all data-quality categories according to the public data quality bitmasks.
To suppress edge effects in the discrete Fourier transform (DFT) of finite-length strain data, we whiten a longer data segment and then perform the DFT on its central portion~\cite{Holdom:2019bdv,Talbot:2025vth}. Specifically, for the target time series $\tilde{d}_I$ of duration $T$, we whiten a segment of duration $3T$, with the target interval placed exactly in the middle.
Then, we employ an iterative notching procedure to the buffered $3T$ segment, following the instrumental line catalogs provided by the \ac{LVK} Collaboration~\cite{gwosc_lines_catalogs}, to mitigate lines of varying strengths~\cite{Ren:2021xbe}. In each iteration, the \ac{PSD} is estimated using only the central $T$ segment to avoid edge artifacts and then is used to whiten the entire $3T$ data stretch. We subsequently analyze the frequency spectrum of the central $T$ segment of the whitened data with squared windowing. Frequency bins within candidate line regions that exceed a predefined amplitude threshold are identified, and corresponding \ac{ZPK} notch filters are constructed and applied to the entire $3T$ segment using zero-phase filtering. After each iteration, $1$\,s of data at both edges is dropped to eliminate filter transients. Upon convergence, we linearly interpolate the \ac{PSD} over a $\pm 2$ Hz interval around the notched frequencies and exclude the central $\pm 0.3$ Hz bins from the likelihood evaluation.

Since the high-\ac{SNR} ringdown events considered in this study are dominated by the two LIGO detectors, we restrict our attention to a network consisting solely of these two detectors. Taking the Hanford detector as the reference, the combined data in Eq.~(\ref{eq:combdata}) can then be written as
\begin{equation}\label{eq:dHL}
    \tilde{d}_j = \tilde{P}_{j} \left( \frac{d_{H,j}}{\tilde{P}_{H,j}}+ A_{HL} \ee^{\ii (\phi_{HL,0}-2\pi f_j \Delta t_{HL})}\frac{d_{L,j}}{\tilde{P}_{L,j}} \right) \,,
\end{equation}
where the combined PSD is $\tilde{P}_j=(\tilde{P}_{H,j}^{-1}+A^2_{HL}\tilde{P}_{L,j}^{-1})^{-1}$.
With only three detector-response parameters $A_{HL}, \phi_{HL,0}$, and $\Delta t_{HL}$, we adopt a minimal approach that does not require detailed sky location information. Specifically,
the arrival time lag $\Delta t_{HL}$ is set to the best-fit value $\Delta t_{HL,0}$ from the main event analysis.
For the remaining two parameters, we note that
the two LIGO detectors are rotated nearly 90 degrees relative to each other, so typical values of $A_{HL}$ and $\phi_{HL,0}$ fall within narrow ranges: $A_{HL}\approx 1$ and $\phi_{HL,0}\approx \pi$.  This motivated our initial choice of priors in Ref.~\cite{Ren:2021xbe}: $A_{HL}=1$ and $\phi_{HL,0}\in [\pi/2,3\pi/2]$. Nonetheless, as the detectors are not perfectly aligned, these parameters can deviate significantly for certain source positions. To account for this possibility, we relax the prior for $\phi_{HL,0}$ to encompass a full $2\pi$ range here, ensuring a coherent combination of signals on two detectors.
We keep $A_{HL}=1$ because the two detectors have comparable \acp{SNR} for the three events considered. As shown later in injection tests, this minimal treatment of the detector-response parameters is generally efficient.

For the effective search template $\tilde{h}_j$ in Eq.~(\ref{eq:combdata}), since it remains close to the original signal waveform as noted earlier, we continue to use the simple \ac{UniEw} model in Eq.~(\ref{eq:uniformQNM}) as the leading-order description of \acp{QNM}, with the detector response now included.
The \ac{UniEw} model is parametrized by seven parameters:
\begin{equation}\label{eq:UniEwparams}
    \{\Delta f, q_0, A^\prime, \tau, f_{\rm min}, f_{\rm max}, T\} \, ,
\end{equation}
which correspond to the frequency spacing, relative offset, overall amplitude, damping time, minimum and maximum central frequencies of the \acp{QNM}, and time duration of the echo signal, respectively. To ensure that each \ac{QNM} is distinct and well separated, we impose a cutoff on the spectral width of each mode defined as $f_\textrm{cut}=\min\{\max[ \sqrt{6}/(\pi\tau),3/T],\Delta f/2\}$. This cutoff is bounded from above by $\Delta f/2$ to prevent overlap between adjacent modes. The lower bound is set to $\max[ \sqrt{6}/(\pi\tau),3/T]$, which ensures that the truncation occurs at approximately $20\%$ of the peak height in the high-resolution regime, or retains at least six frequency bins in the lower-resolution limit.

The parameter space of the \ac{UniEw} model is motivated by the expected echo signal characteristics from \acp{UCO}.
The frequency spacing $\Delta f$ is related to the time delay $t_d$ by $\Delta f\approx 1/t_d$.
For a \ac{UCO} whose reflective surface lies a Planck-scale distance outside the would-be horizon---as may arise from quantum-gravity corrections---we parametrize the separation as
$\ln\epsilon\approx -\eta \ln (M/\ell_{\mathrm{Pl}})$ with $\eta\sim \mathcal{O}(1)$ and $\ell_{\mathrm{Pl}}$ the Planck length. Using Eq.~(\ref{eq:td}), this gives
\begin{equation}\label{eq:Deltaf}
M \Delta f \approx \frac{\bar{R}}{\eta},\quad
\bar{R}=\frac{0.00547}{1+(1-\chi^2)^{-1/2}}\,.
\end{equation}
where $M$ is the detector-frame mass for the remnant \ac{UCO}. For typical \ac{CBC} remnants with $M\sim50M_\odot$ and $\chi\sim 0.7$, the echo time delay is of order $t_d\sim \mathcal{O}(0.1)$s. We therefore take $\eta\in [1,4]$ to cover a broad range of $\Delta f$ values in our search. The relative offset parameter $q_0$ ranging from $0$ to $1$ controls the relative shift of the \acp{QNM} within the frequency band. The upper bound of frequency band $f_{\rm max}$ is tied to the $l=m=2$ fundamental \ac{QNM} frequency of a Kerr \ac{BH} with the same mass and spin, which is well approximated by the analytical fit~\cite{Berti:2005ys,Berti:2009kk}:
\begin{equation}\label{eq:fRD}
   M f_{\mathrm{RD}} \approx \frac{1.5251-1.1568(1-\chi)^{0.1292}}{2\pi} \, .
\end{equation}
Above $f_{\rm RD}$,
as $\mathcal{R}_{\rm eff}(f)$ becomes significantly less than 1, Eq.~(\ref{eq:fntaun}) no longer applies, and the high-frequency \acp{QNM} of \ac{UCO} have significantly shorter damping times.
These short-lived modes give rise to the first few echo wavelets~\cite{Rosato:2025byu} and are better suited to time-domain searches that target burstlike signals~\cite{Tsang:2018uie,Tsang:2019zra,Salemi:2019uea,Conklin:2021cbc,Miani:2023mgl}.
Conservatively, we set $f_{\rm max} < 1.1 f_{\rm RD}$.
The lower bound $f_{\rm min}>50\mathrm{Hz}$ is determined by the detector sensitivity. To ensure a sufficient number of modes within the frequency band, we require $f_{\rm max}-f_{\rm min}>10\Delta f$. The mode width $1/\tau$ is required to satisfy $1/T < 1/\tau < \Delta f$, ensuring that the modes are both frequency-resolved and well separated. Since the signal amplitude of interest in a Bayesian search is not expected to greatly exceed the typical noise level $\langle |n_j|\rangle$
in the frequency band, we choose the prior bounds $A^\prime_{\rm min}=\langle |n_j|\rangle/100$ and $A^\prime_{\rm max}=10\langle |n_j|\rangle$ to ensure convergence. Here $\langle |n_j|\rangle$ denotes the typical frequency-domain strain amplitude in the search band, computed directly from the conditioned data used in the analysis by averaging $|d_{I,j}|$ over frequency bins and over the two LIGO detectors. For the data used here, this quantity
is empirically about
$1.2\langle\tilde{P}_I\rangle^{1/2}$  within the frequency band considered, which
serves as a reference scale for setting a sufficiently broad prior on $A^\prime$.
To resolve long-lived \acp{QNM}, the strain data must contain a sufficiently long sequence of echo wavelets. For illustration, we take this number to be of order  $\mathcal{O}(100)$, corresponding to duration $T\in [100/\Delta f_{\rm min},200/\Delta f_{\rm min}]$ for two benchmark values.

\begin{table}[h]
    \renewcommand{\arraystretch}{1.2}
    \setlength{\tabcolsep}{5pt}
\begin{center}
\begin{tabular}{l@{\hspace{2em}}l}
\hline\hline
&
\\[-3mm]
Parameters & Priors and fixed (scanned) values
\\
&
\\[-3.5mm]
\hline
$\Delta f$ & Uniform in $[\bar{R}/4, \bar{R}]\times c^3/(GM)$
\\
$q_0$ & Uniform in $[0,1]$
\\
$A^\prime$ & Uniform in $[\langle |n_j| \rangle/100\,,10\langle |n_j| \rangle]$
\\
$1/\tau$ & Log-uniform in $[1/T, \Delta f_{\rm max}]$
\\
$f_\textrm{min}$, $f_{\rm max}$ & Uniform in $[50\mathrm{Hz}, 1.1 f_{\rm RD}]$
\\
 & With $f_{\rm max}-f_{\rm min}>10\Delta f$
\\
$\phi_{HL,0}$ & Uniform in $[-\pi/2,3\pi/2]$
\\
\hline
$A_{HL}$ & $1$
\\
$\Delta t_{HL}$ & $\Delta t_{HL,0}$
\\
\hline
$T$ & $400GM/(\bar{R}c^3)$ and $800GM/(\bar{R}c^3)$
\\
\hline\hline

\end{tabular}
\caption{Parameter settings for the LVK data search. The search parameters $\Delta f$, $q_0$, $A^\prime$, $\tau$, $f_\textrm{min}$, $f_\textrm{max}$, $\phi_{HL,0}$ are listed with their priors.
The prior on $A^\prime$ is defined in terms of the typical frequency-domain noise level
$\langle |n_j|\rangle$(see text for details).
$\Delta f_{\rm max}$ denotes the upper bound of the prior range of $\Delta f$. The derived quantities $\bar{R}$ and $f_\textrm{RD}$ are given by Eqs.~(\ref{eq:Deltaf}) and (\ref{eq:fRD}), respectively. $M$ and $\chi$ denote the best-fit remnant mass (detector frame) and spin obtained from the main event analysis.
$A_{HL},\Delta t_{HL}$ are fixed parameters. Variation of the time duration $T$ is implemented by scanning over two benchmark values.
All quantities in this table are expressed in SI units, with $c$ and $G$ denoting the speed of light and the gravitational constant, respectively.}
\label{tab:parameter}
\end{center}
\end{table}

Table~\ref{tab:parameter} summarizes the parameter settings used in our LVK data search. The priors for the six intrinsic search parameters $\Delta f$, $q_0$, $A^\prime$, $\tau$, $f_{\rm min}$ and $f_{\rm max}$ closely follow those adopted in our previous Gaussian-noise analysis~\cite{Wu:2023wfv}. For the detector-response parameters, we use the same settings as in our previous O1 study~\cite{Ren:2021xbe}, except that the prior on the relative phase $\phi_{HL,0}$ has been relaxed to cover special sky locations.

We employ the \textsc{bilby} \cite{Ashton:2018jfp} package to estimate the Bayes factor and the posterior distributions with its integrated \textsc{dynesty} nested sampler \cite{Speagle:2019ivv}. To balance accuracy and computational efficiency, we employ a sampler configuration broadly inspired by Refs.~\cite{Ren:2021xbe,Wu:2023wfv}, with minor empirical adjustments. In practice, we use $\mathrm{sample}=\text{rwalk}$, $n_{\rm live}=1000$, ${\rm walks}=100$, and ${\rm maxmcmc}=10000$, which provide stable convergence and sufficient sampling accuracy for the present analysis.

\section{Performance validation on real data}\label{sec:inject}

In this section, we validate the performance of the search algorithm using the LVK data. Given that the O1 data contain stronger non-Gaussian artifacts compared to later runs, we use the detector noise around GW150914 for this test. To compare with the Gaussian-noise results of Ref.~\cite{Wu:2023wfv}, we vary the time duration over $T \in \{48.6, 36.5, 24.3, 12.1, 4.9\}$\,s. The search and fixed parameter settings remain identical to those listed in Table~\ref{tab:parameter}.

We first evaluate the performance of the algorithm on background noise. Independent noise realizations are generated using the time-slide method \cite{Capano:2017fia,DiMarco:2023jqq}. For GW150914, we select $25$ consecutive strain data segments beginning $250$\,s prior to the event time. Each segment has a duration $T$ and is indexed by $k=0, 1, \dots, 24$, with segments containing glitches excluded based on data-quality checks. Noise realizations are constructed by combining segments from the two detectors with differing indices ($k_H \neq k_L$). This procedure introduces a time shift $|k_{H}-k_{L}|T$ significantly larger than the light travel time between the two detectors, ensuring the statistical independence of the background samples.
For the notching procedure, following Ref.~\cite{Ren:2021xbe}, we target instrumental lines originating from 60\,Hz power-mains harmonics and the \ac{OMC} length dither. For each detector $I$, we notch listed lines whose normalized strain amplitude $d_{I,j} / \smash{\tilde{P}_{I,j}}^{1/2}$ exceeds a threshold of 6.

\begin{figure}[htbp]
	\centering
	\includegraphics[width=15cm]{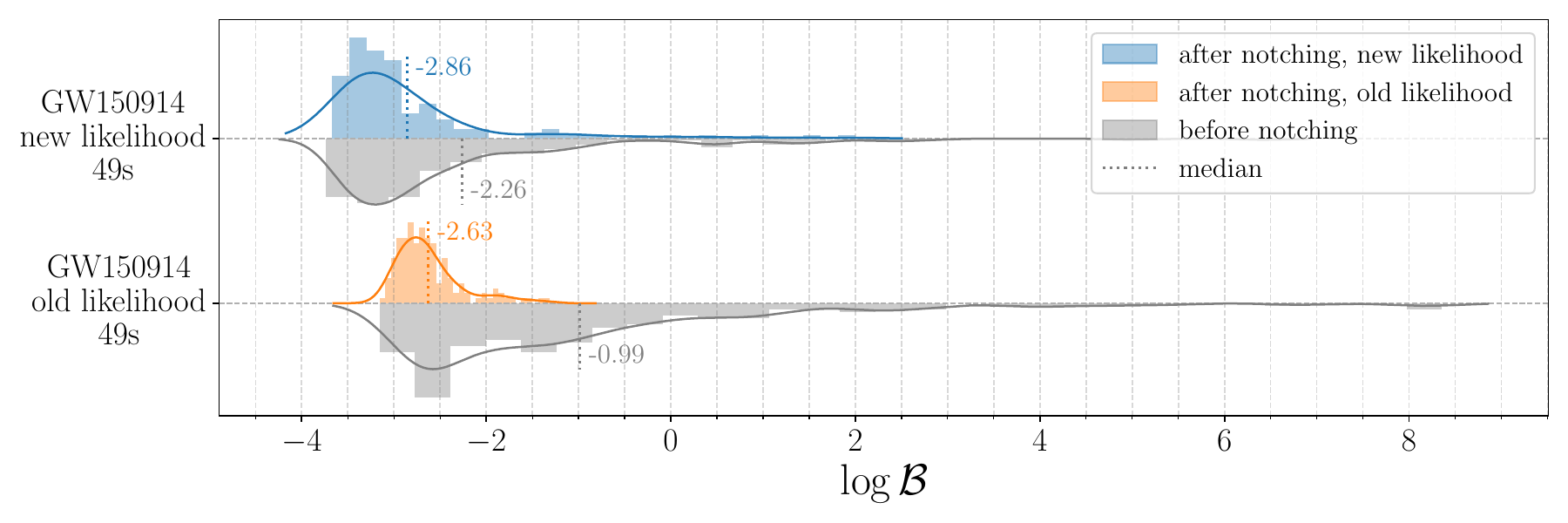}
	\caption{\label{fig:notchcompare}{Comparison of background log Bayes factor distributions with and without notch filtering for detector noise around GW150914. The blue and orange histograms show the notched distributions obtained with the new and old likelihoods, respectively; gray histograms correspond to the results before notching. The vertical dotted lines mark the distribution medians. Each histogram is constructed from $150$ independent noise realizations.}}
\end{figure}

Figure~\ref{fig:notchcompare} compares the log Bayes factor distributions obtained from real data around GW150914 with and without notch filtering.
The longest duration case is chosen to illustrate the effect of notching, as this duration is representative of the actual search.
After notch filtering, the background distribution (orange histograms) obtained with the old likelihood shows a significant reduction in the number of high log Bayes factor outliers compared to the non-notched case (gray histograms), in agreement with our earlier findings~\cite{Ren:2021xbe}. In contrast, when employing the new likelihood (blue histograms), the difference between the notched and non-notched cases becomes less pronounced, demonstrating that the coherent phase combination inherent to the new likelihood already suppresses most line-induced contamination. Nevertheless, notch filtering still removes a few residual outliers for this case; therefore, to ensure robustness, we retain the notch step in all analyses of real data.
Overall, for both likelihoods, background distributions of the log Bayes factor after notching remain relatively stable for different time durations considered here and match the corresponding Gaussian results~\cite{Wu:2023wfv}.

Next, we perform injection tests on real data. For comparison with the Gaussian-noise results,
we adopt the representative benchmark echo waveform shown in Figure~\ref{fig:waveform}, which has also been studied in Ref.~\cite{Wu:2023wfv}.
This waveform is injected into the same $150$ independent noise realizations used in the background study. The injection amplitude is adjusted to yield an optimal network \ac{SNR} of approximately 16 for a duration of $T=49$s. In this configuration, the signal contains approximately $200$ echo wavelets, and the ringdown component has an \ac{SNR} of about $6$.
For the injection, we set the sky location to right ascension $\alpha=1.68$\,rad, declination $\delta=-1.21$\,rad and \ac{GMST}$=9.383$\,hour, yielding $\phi_{HL,0}\sim 0$ and $\Delta t_{HL,0}=7.4$\,ms.
This corresponds to the special scenario in which the relative phase $\phi_{HL,0}$ differs significantly from $\pi$---the value expected if the two LIGO detectors were oriented exactly 90 degrees apart.

\begin{figure}[htbp]
	\centering
	\includegraphics[width=16cm]{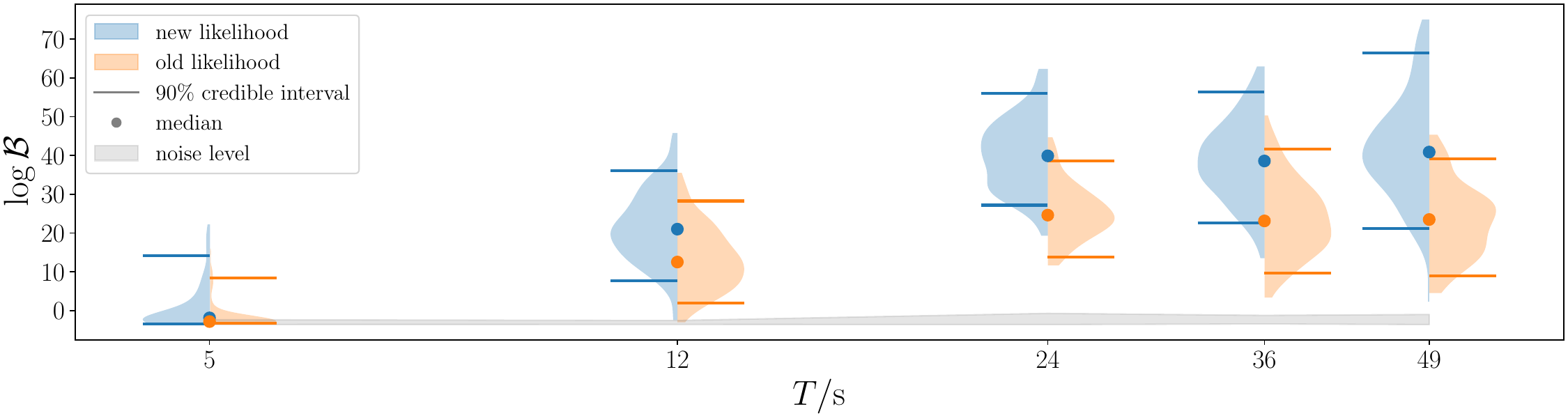}
	\includegraphics[width=16cm]{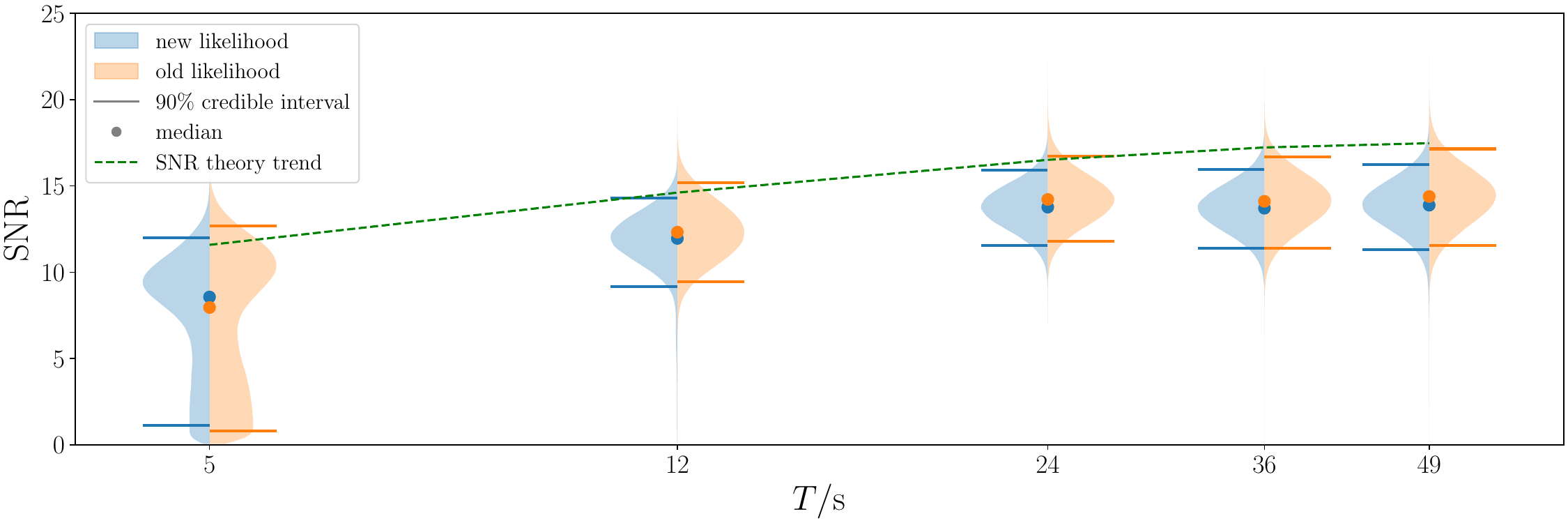}
        \includegraphics[width=16cm]{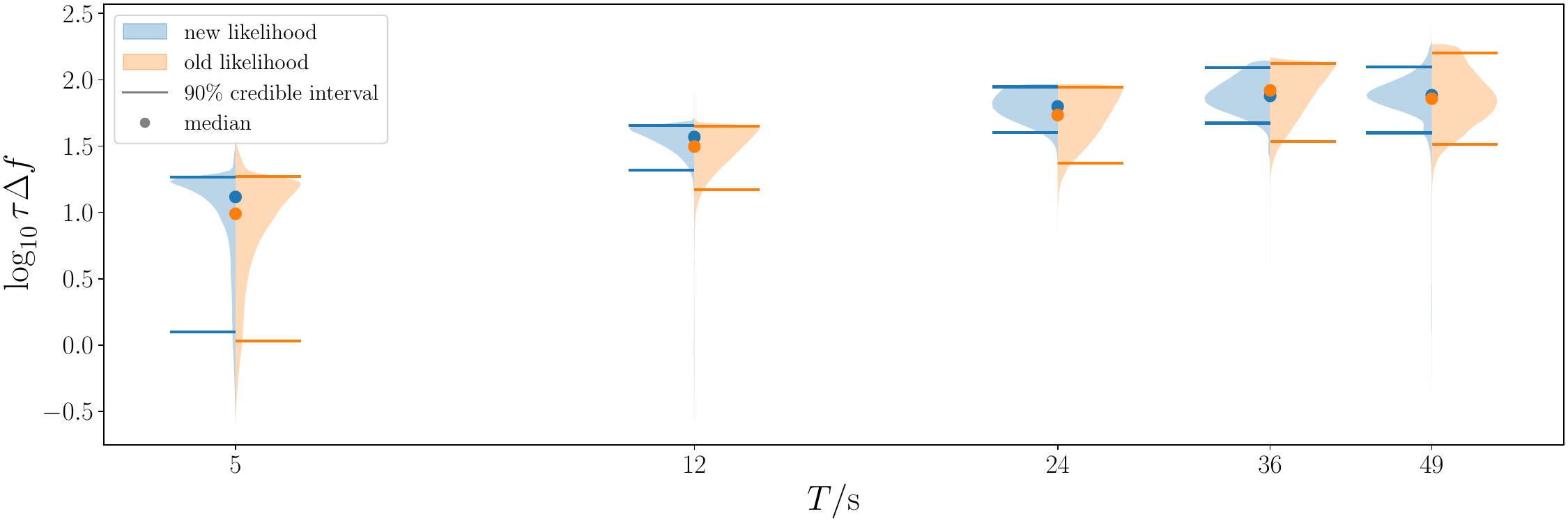}
	\caption{\label{fig:violininj}{The log Bayes factor distributions (top), the overall posterior distributions of the inferred network \ac{SNR} (middle) and the combination $\tau \Delta f$ (bottom) for the injection, as a function of the time duration $T$. For all panels, the upper and lower bars represent the symmetric 90\% credible intervals, and the dots denote the median values. In the top panel, the gray band represents the 90\% credible interval of the noise distributions. For the middle ones, the green dashed lines denote the theoretical prediction of the \ac{SNR}.
 }}
\end{figure}

\begin{figure}[htbp]
	\centering
	\includegraphics[width=6cm]{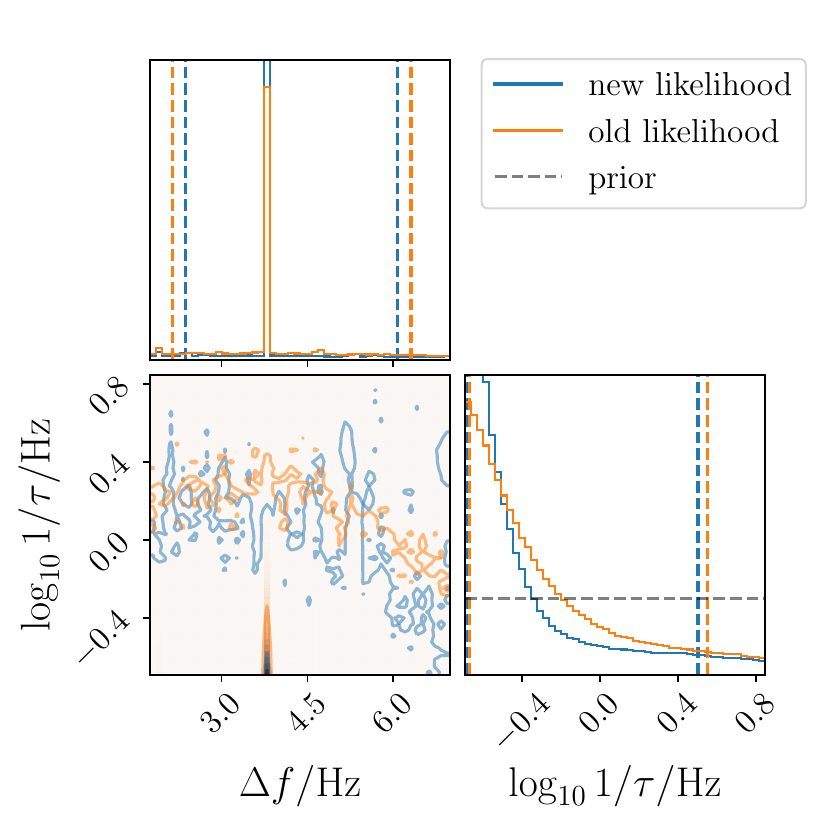} \quad\hspace{-0.3cm}
    \includegraphics[width=6cm]{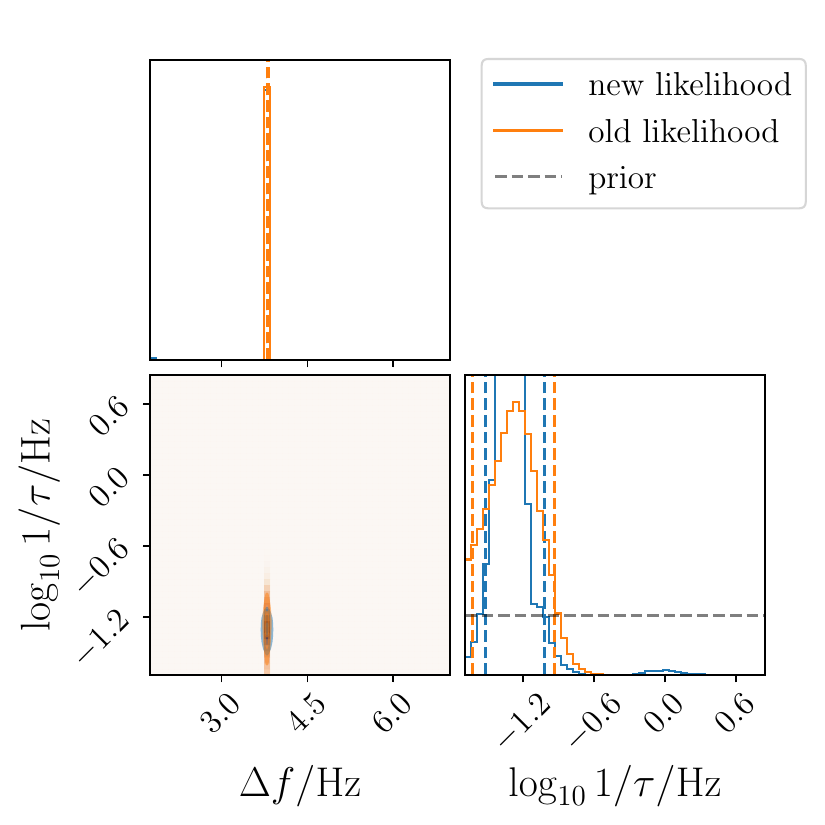} \hspace{-0.5cm}\\
   \includegraphics[width=6cm]{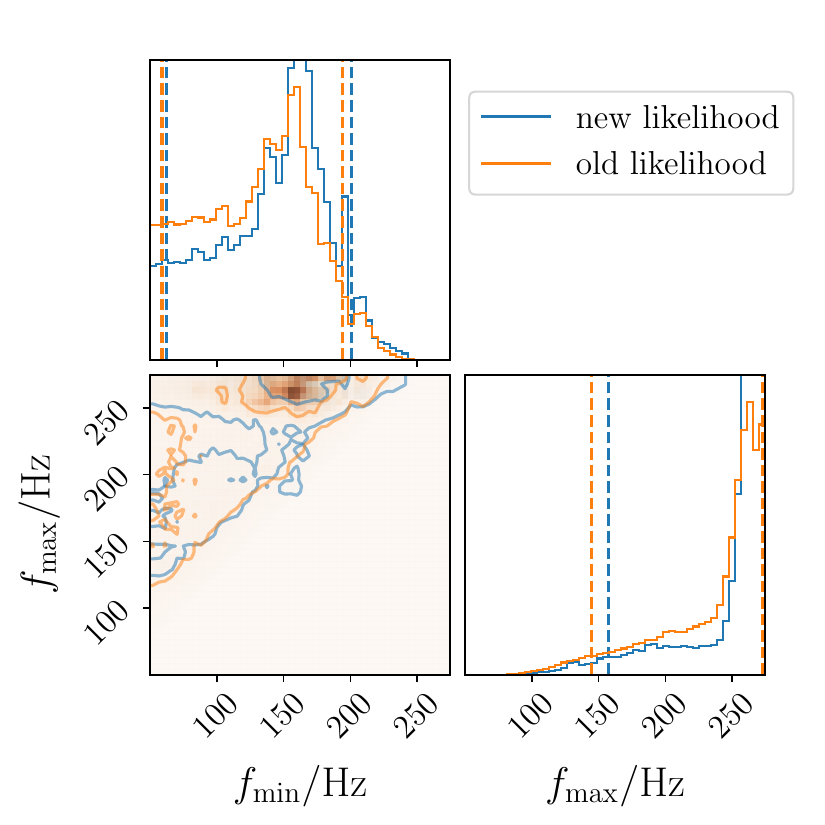}\quad\hspace{-0.2cm}
    \includegraphics[width=6cm]{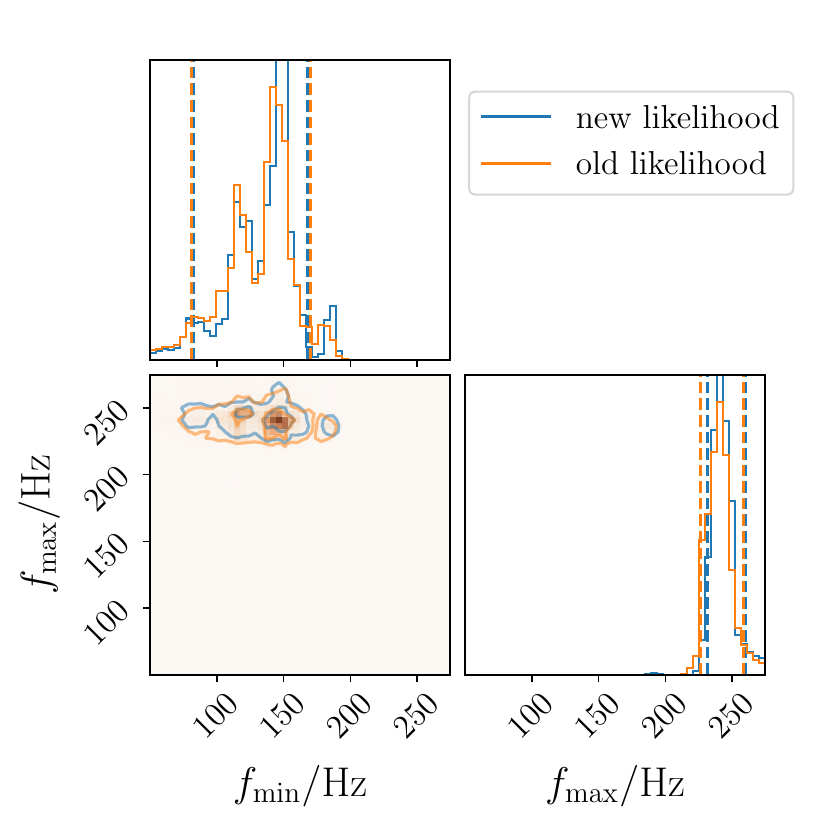}
	\caption{\label{fig:posterior1}{Corner plots for the overall posterior distributions for the spacing $\Delta f$, the width $1/\tau$ and the frequency range $f_{\rm min}, f_{\rm max}$ for the injection with $T=5$\,s (left) and $49$\,s (right). The blue and orange are for new and old likelihoods, respectively. The contours in the bottom-left panels denote the $1\sigma$ and $2\sigma$ ranges of the 2d posteriors.}}
\end{figure}

\begin{table}[htbp]
    \renewcommand{\arraystretch}{1.2}
    \setlength{\tabcolsep}{2pt}
\centering
\begin{tabular}{ccccccccccc}
\hline\hline

likelihood & $T/\mathrm{s}$ & $\log \mathcal{B}$ & $\Delta f/\mathrm{Hz}$ & $\log_{10}(1/\tau/\mathrm{Hz})$ & $q_0$ & $A^\prime/\langle |n_j|\rangle$ & $f_{\rm min}/\mathrm{Hz}$ & $f_{\rm max}/\mathrm{Hz}$ & $\phi_{HL,0}$ & \ac{SNR}\\
\hline
\multirow{5}{*}{new}
& 5 & $-1.8^{+16.0}_{-1.5}$ & $3.8^{+2.3}_{-1.4}$ & $-0.5^{+1.0}_{-0.1}$ & $0.88^{+0.10}_{-0.84}$ & $0.9^{+0.6}_{-0.8}$ & $150^{+50}_{-90}$ & $260.2^{+12.9}_{-103.1}$ & $0.3^{+3.4}_{-1.1}$ & $9^{+3}_{-7}$ \\
& 12 & $21.0^{+15.1}_{-13.2}$ & $3.812^{+0.002}_{-0.002}$ & $-0.99^{+0.25}_{-0.09}$ & $0.95^{+0.03}_{-0.05}$ & $1.4^{+0.5}_{-0.5}$ & $150^{+40}_{-60}$ & $252.5^{+18.0}_{-15.7}$ & $0.1^{+0.5}_{-0.5}$ & $12^{+2}_{-3}$ \\
& 24 & $39.9^{+16.1}_{-12.7}$ & $3.8120^{+0.0008}_{-0.0008}$ & $-1.2^{+0.2}_{-0.1}$ & $0.98^{+0.01}_{-0.01}$ & $1.5^{+0.4}_{-0.4}$ & $140^{+40}_{-30}$ & $249.8^{+14.1}_{-13.5}$ & $0.0^{+0.3}_{-0.3}$ & $14^{+2}_{-2}$ \\
& 36 & $38.6^{+17.7}_{-15.9}$ & $3.8119^{+0.0006}_{-0.0007}$ & $-1.3^{+0.2}_{-0.2}$ & $0.986^{+0.008}_{-0.009}$ & $1.3^{+0.6}_{-0.4}$ & $130^{+40}_{-50}$ & $244.7^{+20.0}_{-13.5}$ & $0.1^{+0.4}_{-0.3}$ & $14^{+2}_{-2}$ \\
& 49 & $40.9^{+25.6}_{-19.7}$ & $3.8119^{+0.0006}_{-0.0006}$ & $-1.3^{+0.3}_{-0.2}$ & $0.982^{+0.007}_{-0.009}$ & $1.2^{+0.4}_{-0.4}$ & $140^{+30}_{-60}$ & $242.0^{+18.1}_{-10.4}$ & $-0.0^{+0.5}_{-0.4}$ & $14^{+2}_{-3}$ \\
\hline
\multirow{5}{*}{old}
& 5 & $-2.8^{+11.2}_{-0.4}$ & $3.8^{+2.5}_{-1.7}$ & $-0.4^{+1.0}_{-0.3}$ & $0.8^{+0.2}_{-0.7}$ & $0.7^{+0.8}_{-0.6}$ & $130^{+60}_{-80}$ & $250^{+20}_{-110}$ & $0.4^{+3.6}_{-1.4}$ & $8^{+5}_{-7}$ \\
& 12 & $12.5^{+15.7}_{-10.5}$ & $3.812^{+0.003}_{-0.003}$ & $-0.9^{+0.3}_{-0.2}$ & $0.95^{+0.04}_{-0.09}$ & $1.3^{+0.6}_{-0.5}$ & $160^{+30}_{-70}$ & $251.4^{+19.5}_{-15.1}$ & $0.1^{+0.5}_{-0.5}$ & $12^{+3}_{-3}$ \\
& 24 & $24.6^{+14.0}_{-10.7}$ & $3.812^{+0.001}_{-0.002}$ & $-1.2^{+0.4}_{-0.2}$ & $0.98^{+0.02}_{-0.96}$ & $1.4^{+0.6}_{-0.5}$ & $140^{+30}_{-50}$ & $249.0^{+17.0}_{-16.0}$ & $0.0^{+0.4}_{-0.4}$ & $14^{+2}_{-2}$ \\
& 36 & $23.1^{+18.5}_{-13.4}$ & $3.8118^{+0.0009}_{-0.0008}$ & $-1.3^{+0.4}_{-0.2}$ & $0.986^{+0.010}_{-0.018}$ & $1.4^{+0.6}_{-0.5}$ & $130^{+40}_{-50}$ & $243.2^{+22.0}_{-14.0}$ & $0.0^{+0.4}_{-0.4}$ & $14^{+3}_{-3}$ \\
& 49 & $23.5^{+15.7}_{-14.6}$ & $3.8119^{+0.0010}_{-0.0008}$ & $-1.3^{+0.3}_{-0.3}$ & $0.983^{+0.009}_{-0.016}$ & $1.2^{+0.6}_{-0.4}$ & $140^{+30}_{-60}$ & $240.2^{+18.3}_{-13.9}$ & $-0.0^{+0.5}_{-0.4}$ & $14^{+3}_{-3}$ \\
\hline\hline

\end{tabular}
\caption{Search results for the injection tests.
The ``$\log \mathcal{B}$'' column reports the median and symmetric 90\% intervals of the log Bayes factors obtained from the distribution of the 150 independent noise realizations. The remaining columns present the median values and 90\% credible regions \cite{LIGOScientific:2013yzb} for the search parameters and the inferred network \ac{SNR}, derived from the overall posterior distributions.
(See the text for further discussion of the meaning of the inferred \ac{UniEw} parameters and their relation to the injected signal.)}
\label{tab:injectResult}
\end{table}

Figures~\ref{fig:violininj} and \ref{fig:posterior1}, together with Table~\ref{tab:injectResult}, summarize the Bayesian search results for the injection tests as a function of signal duration $T$.
To capture the average inference behavior in real detector noise, we show overall posterior distributions obtained from $150$ noise realizations. These overall posteriors are constructed by concatenating posterior samples from all noise realizations with equal weights~\cite{Breschi:2022ens,Wu:2023wfv}.

In Figure~\ref{fig:violininj}, the top panel shows the distributions of log Bayes factors for the injections (colored regions) together with the noise background (gray band). In the short-duration regime ($T\lesssim 12$s), the log Bayes factors of the injected signals are only marginally higher than those of the noise, indicating limited detection significance. As $T$ increases, the log Bayes factors obtained with the new likelihood increase steadily and then level off at values around $40$ for $T\gtrsim 24$\,s, demonstrating a strong and stable detection. By contrast, the log Bayes factors obtained with the old likelihood initially increase with $T$ but exhibit a mild decline at larger durations, although a positive detection is still achieved for $T\gtrsim 24$s. This behavior arises because the old likelihood relies solely on amplitude information. As $T$ increases, the signal is distributed across a larger number of frequency bins, leading to a reduction of the \ac{SNR} per bin. In contrast, the new likelihood consistently yields higher log Bayes factors, with the improvement becoming more pronounced at larger $T$. This enhancement stems from the coherent integration of signals across multiple frequency bins, which effectively leverages phase information to boost the detection significance.
The middle panel presents the overall posterior distributions of the inferred \ac{SNR} for the injections, where the \ac{SNR} is computed using the search template with the overall posterior distributions of the search parameters.
The theoretical prediction (green dashed line) is computed directly from the injected waveform over a broader frequency range (i.e. $50\mathrm{Hz}<f\lesssim 274{\rm Hz}$) and therefore yields higher values than the inferred \ac{SNR} from the search. The inferred \ac{SNR} follows a similar trend as a function of $T$, indicating that the search template successfully captures the dominant features of the injected signal.
The bottom panel displays the overall posterior distributions of the combination $\tau \Delta f$.
Following Eq.~(\ref{eq:fntaun}), this product corresponds to the inverse logarithmic reflectivity, $1/|\ln\mathcal{R}_{\rm eff}(f_n)|\approx \tau_n\Delta f$, from which we can infer the average value of $\mathcal{R}_{\rm eff}(f)$.
The upper limits of the 90\% credible intervals from both likelihoods are constrained by the prior boundary $\tau < T$. As $T$ increases, the posterior distributions become more symmetric, reflecting improved constraints on the average reflectivity $\mathcal{R}_{\mathrm{eff}}(f)$.

For parameter estimation, Figure~\ref{fig:posterior1} presents the corner plots for several key intrinsic parameters, and Table~\ref{tab:injectResult} lists the inferred values for all search parameters.
Overall, the inferred values of the six intrinsic parameters for injections on real data are consistent with those obtained with Gaussian noise~\cite{Wu:2023wfv}, demonstrating the negligible effect of residual non-Gaussian artifacts on both echo detection and parameter estimation.
Because the injected signal is a benchmark waveform rather than the \ac{UniEw} search template, these parameters do not have direct injected counterparts. As noted earlier, however, the \ac{UniEw} template is designed to capture the dominant \acp{QNM} of the realistic spectrum, with its parameters providing an estimate of the average mode properties. For this injected signal, the mode spacing varies only slightly within the inferred frequency range, and the inferred spacing $\Delta f$ of \ac{UniEw}
is consistent with the expected nearly uniform comblike spacing of the dominant modes in the searched band, with a characteristic scale approximately given by $1/t_d$.
The inferred upper frequency bound $f_{\rm max}$ approaches the prior limit $1.1f_{\rm RD}\approx 274$\,Hz, while the lower bound $f_{\rm min}$ clusters around $140$\,Hz.
This is consistent with the frequency band that contains the SNR-dominant \acp{QNM} (see Figure~\ref{fig:waveform}), and matches the fact that the inferred \ac{SNR} is only slightly below the injected value.
Using the posterior median values of $\tau$ and $\Delta f$, we obtain an effective estimate $1/|\ln\mathcal{R}_{\rm eff}|\approx \tau\Delta f\approx 76$ within the searched band, which is broadly consistent with the theoretical expectation discussed in Ref.~\cite{Wu:2023wfv}.
For the detector response, the posterior of the only free parameter $\phi_{HL,0}$ successfully recovers the injected value, validating the relaxation of its prior to the full $2\pi$ range.
Together, these results validate that the simple \ac{UniEw} template can effectively capture dominant long-lived-\ac{QNM} content of the injected echo signal in real detector noise and provide useful effective parameter estimates.

In summary, the results for background noise and injection tests on real detector noise align with expectations from idealized studies, validating that the proposed search pipeline performs robustly under realistic conditions.
This validation provides a solid foundation for applying the method to real gravitational-wave data.

\section{Results and Discussions}\label{sec:result}
Having validated the methodology with O1 data, we now apply the proposed search to confirmed CBC events that exhibit high ringdown \ac{SNR}.
To allow for comparison with our earlier study, we retain GW150914 and include two recent events from the O4 observing run: GW231226 and GW250114. These events were selected for their high network \ac{SNR} and detection by both LIGO detectors. All three remnants have masses near $70\,M_{\odot}$, placing the frequency band of interest between $\sim 50$\,Hz to $\sim 250\,$Hz.
The O4 data offer two key improvements over O1 for searching long-lived echo \acp{QNM}. First, the noise floor $\langle|n_j|\rangle$ in the relevant frequency band is reduced by a factor of 2--3. Second, the data contain significantly fewer instrumental lines. This reduction minimizes the number of notched lines, yielding a cleaner spectrum after filtering. Below we present the search results for the three selected events. We begin with an analysis of the noise background using data away from the event in the segment, followed by the results of the postmerger echo search.

\subsection{Noise background}
\begin{table}[h]
    \renewcommand{\arraystretch}{1.2}
    \setlength{\tabcolsep}{2pt}
\begin{center}
\begin{tabular}{llll}
\hline\hline
Parameters & GW150914 & GW231226 & GW250114
\\
\hline
\multicolumn{4}{l}{\hspace{0.5em}\textbf{Search Configuration}} \\
$\Delta f$ & Uniform in $[1.75,6.99]$\,Hz & Uniform in $[1.38,5.52]$\,Hz & Uniform in $[1.72,6.90]$\,Hz
\\
$q_0$ & Uniform in $[0,1]$ & Uniform in $[0,1]$ & Uniform in $[0,1]$
\\
$A^\prime$ & Uniform in $[\langle |n_j|\rangle/100\,,10\langle |n_j|\rangle]$ & Uniform in $[\langle |n_j|\rangle/100\,,10\langle |n_j|\rangle]$ & Uniform in $[\langle |n_j|\rangle/100\,,10\langle |n_j|\rangle]$
\\
$1/\tau$ & \makecell[l]{Log-uniform in $[0.0175, 6.99]$\,Hz \\ and $[0.0087, 6.99]$\,Hz} & \makecell[l]{Log-uniform in $[0.0138, 5.52]$\,Hz \\ and $[0.0069, 5.52]$\,Hz} & Log-uniform in $[0.0172, 6.90]$\,Hz
\\
$f_\textrm{min}$, $f_{\rm max}$ & \makecell[l]{Uniform in $[50,274.5]$\,Hz \\ With $f_{\rm max}-f_{\rm min}>10\Delta f$} & \makecell[l]{Uniform in $[50,216.7]$\,Hz \\ With $f_{\rm max}-f_{\rm min}>10\Delta f$} & \makecell[l]{Uniform in $[50,274.8]$\,Hz \\ With $f_{\rm max}-f_{\rm min}>10\Delta f$}
\\
$\phi_{HL,0}$ & Uniform in $[-\pi/2,3\pi/2]$ & Uniform in $[-\pi/2,3\pi/2]$ & Uniform in $[-\pi/2,3\pi/2]$
\\
$A_{HL}$ & $1$ & $1$ & $1$
\\
$\Delta t_{HL}$ & $6.9\times 10^{-3}$\,s & $-4.0\times 10^{-3}$\,s & $2.45\times 10^{-3}$\,s
\\
$T$ & $57.2$\,s and $114.4$\,s & $72.4$\,s and $144.9$\,s  & $58.0$\,s
\\
\makecell[l]{start GPS time for \\ postmerger search} & $1126259462.44$\,s & $1387620938.33$\,s & $1420878141.24$\,s
\\
\hline
\multicolumn{4}{l}{\hspace{0.5em}\textbf{Line Notch Settings}} \\
Line origin & \makecell[l]{power mains,\\ \ac{OMC} length dither } & power mains & power mains\\
Threshold & 6 & 6 & 6 \\
\hline
\multicolumn{4}{l}{\hspace{0.5em}\textbf{Time-slide Configuration}} \\
\makecell[l]{segments prior or \\ subsequent time} & $250$\,s & $100$\,s & \makecell[l]{prior time $50$\,s \\ subsequent time $130$\,s} \\
segments number & $25$ and $15$ & $25$ and $14$ & \makecell[l]{prior $2$ segments \\ subsequent $12$ segments}\\
noise realization number & $500$ and $150$ & $500$ and $150$ &$150$\\
\hline\hline

\end{tabular}
\caption{Summary of parameter settings for the real-data search of the three events. The table is organized into three panels: (top) search configuration, including explicit priors for search parameters and values for fixed or scanned settings; (middle) notch-filtering settings, specifying the origin of instrumental lines and the normalized strain amplitude threshold for removal; (bottom) background generation setup, detailing the configurations of the time-slide method.}
\label{tab:parameterevent}
\end{center}
\end{table}

The strain-data preparation follows the same procedure outlined in Sec.~\ref{sec:lvk}, with event-specific parameters listed in Table~\ref{tab:parameterevent}.
For background generation via the time-slide method, we use multiple consecutive preevent segments to build independent noise realizations  for GW150914 and GW231226, where sufficient valid data before the event exist. This yields 500 realizations for the short-duration analysis and 150 for the long-duration analysis. For GW250114, limited preevent data quality requires using segments from both before and after the event, producing only 150 short-duration realizations.
In the notching procedure, only the power-mains harmonics are relevant for GW231226 and GW250114, given the far fewer instrumental lines in O4 compared to O1. The same removal threshold is used for all observing runs. All search parameters follow the general specifications in Table~\ref{tab:parameter}, with event-specific values of $M$, $\chi$, and $\Delta t_{HL}$ taken from the main-event analysis. The postmerger search start time is set after the GPS time of the main event to avoid contamination from the early ringdown signal.

\begin{figure}[htbp]
	\centering
	\includegraphics[width=16cm]{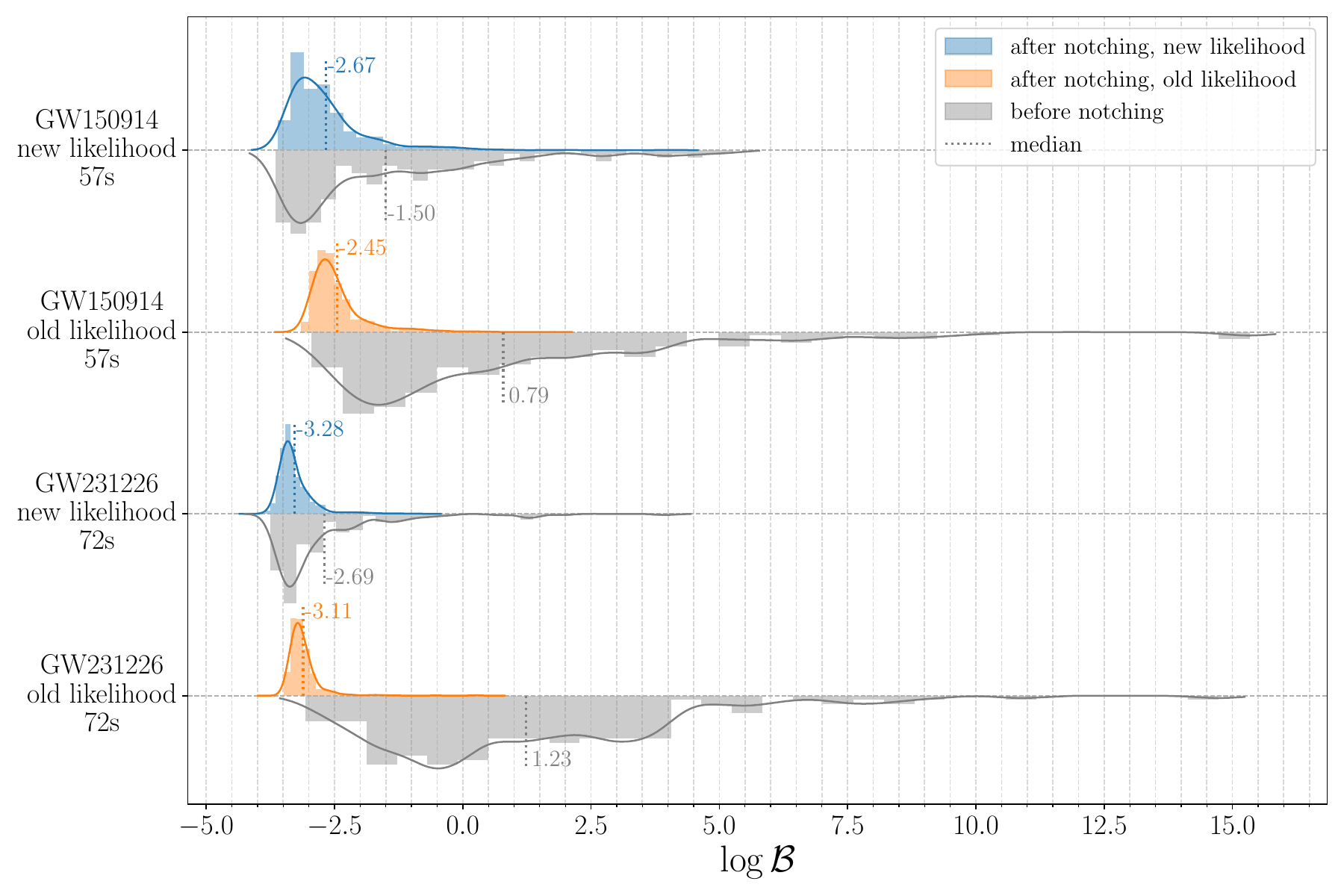}
	\caption{\label{fig:violinnotch}{Comparison of background log Bayes factor distributions with and without notch filtering for detector noise around GW150914 and GW231226. The blue and orange histograms represent the notched distributions obtained with the new and old likelihoods, using $500$ noise realizations, respectively. The gray histograms correspond to the results before notching, using $150$ noise realizations.}}
\end{figure}

Before presenting the noise-background search results, we first demonstrate the robustness of the notching procedure across O1 and O4.
Figure~\ref{fig:violinnotch} compares the distributions of log Bayes factors for the noise background with and without notch filtering for the short-duration case.
For GW150914, the outlier tail in the distribution before notching is longer than in the shorter-duration case of Figure~\ref{fig:notchcompare}, but applying the notch substantially reduces it. For GW231226, outliers before notching arise only from power-mains harmonics, producing a pronounced tail for the old likelihood; after notching, the distribution shifts to more negative values than for GW150914, likely due to the different prior ranges used for O4 events.
For GW250114, which lies in O4b, i.e., the second part of the fourth observing run following O4a and the intervening commissioning break, no dedicated public instrumental-line catalog for that O4b data segment was available at the time of our analysis. We therefore apply the same power-mains notching prescription as for GW231226. In the analyzed band, the only candidate features that exceed our notching threshold are power-mains harmonics, so the practical effect of notching is similar to that for GW231226.
In all cases, longer durations $T$ allow more lines to exceed the threshold and thus require notching. This effect is especially pronounced for GW150914, because O1 data are more heavily contaminated by instrumental lines.

\begin{figure}[htbp]
	\centering
	\includegraphics[width=16cm]{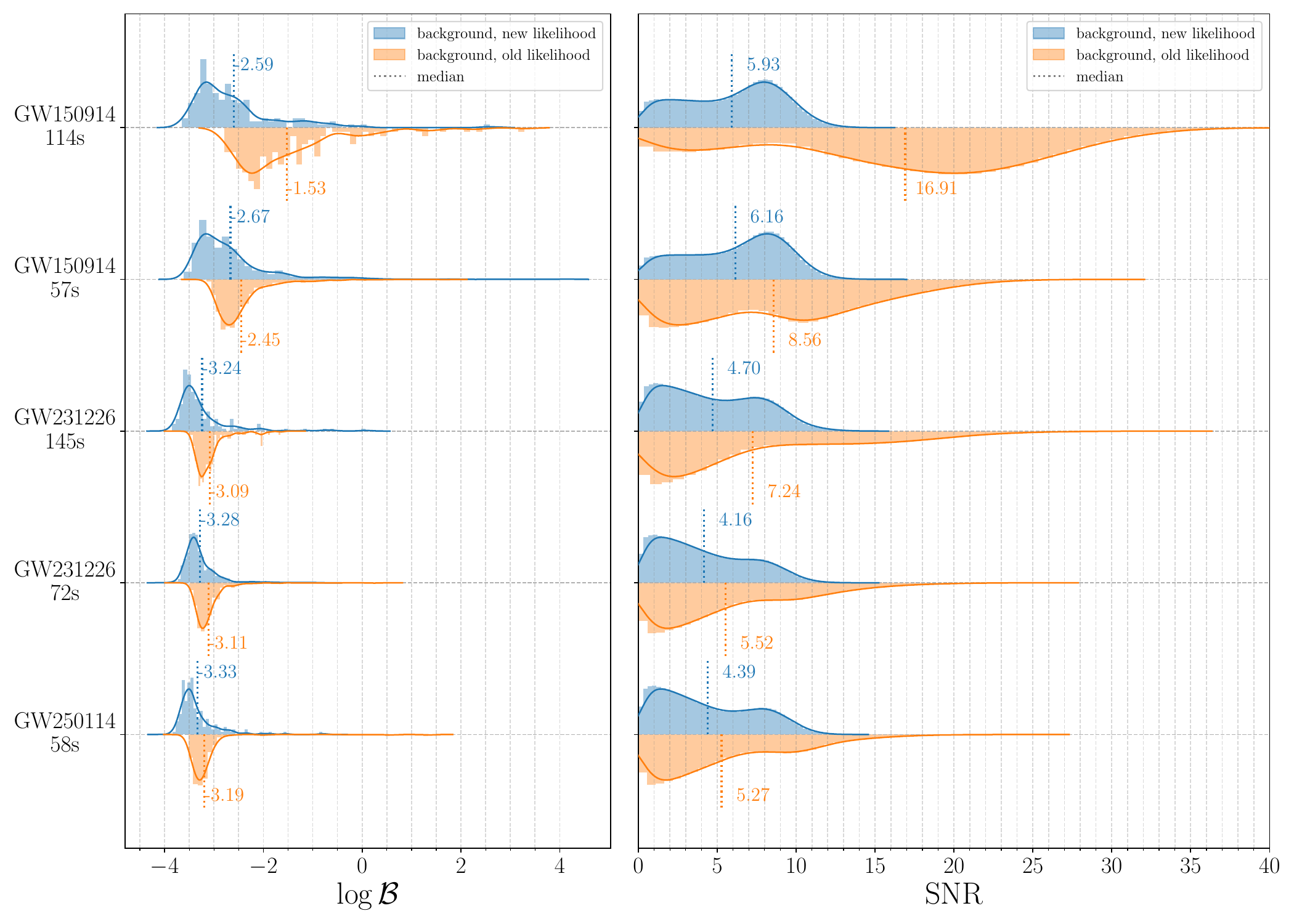}
	\caption{\label{fig:violinbackground}{Background search results for the three events. Left:  the background log Bayes factor distributions. Right: overall posterior distributions of the inferred network \ac{SNR}. Blue and orange denote the new and old likelihood, respectively; vertical dotted lines mark the distribution medians.}}
\end{figure}

Figure~\ref{fig:violinbackground}
summarizes the search results for the noise backgrounds around the three events, using both phase-marginalized likelihood functions. After notching, the background distributions of log Bayes factor  are consistent with stationary Gaussian noise, confirming the effectiveness of the notching procedure. Nevertheless, a tail of noise outliers with high $\log\mathcal{B}$ values persists in some cases, indicating residual non-Gaussian features beyond the obvious instrumental lines.
The overall posterior for the \ac{SNR} varies more from case to case. Its general shape resembles the Gaussian counterpart (see Appendix~\ref{app:gaussian} for details), in which the new likelihood distribution is more stable against variations in duration. The old likelihood, in contrast, develops a longer tail as duration increases---a direct consequence of the larger number of frequency bins included in the search. Additionally, a secondary peak appears at larger \ac{SNR} values due to the influence of non-Gaussian artifacts; this peak is closely linked to the noise outliers that exhibit high $\log\mathcal{B}$ values.

Among all cases examined, we identify three cases with distinct outliers.
For GW150914 with both durations, the new likelihood flags a few noise outliers that correspond to narrow spectral lines whose phase evolution closely resembles that of our target mode. Due to coincidental phase alignment, these lines appear visually similar to the expected signal, give a large likelihood contribution, and yield a best-fit \ac{SNR} around 8, illustrating the improved sensitivity of the algorithm to signal-like structures.
For GW250114, the old likelihood also detects noise outliers that match narrow spectral lines. We observe that these narrow line structures are nonpersistent (absent in other segments/durations) and appear predominantly in a single detector, making them inconsistent with an astrophysical signal.
Interestingly, for GW150914 at longer duration, the old likelihood identifies a distinct, broad, quasiperiodic feature that was not seen in previous studies. Its width is substantially larger than typical lines, producing a large secondary peak in the \ac{SNR} posterior near 20. Its origin requires further investigation. Additional details on these artifacts are provided in Appendix~\ref{app:line}.

\subsection{Postmerger search}

\begin{table}[h]
    \renewcommand{\arraystretch}{1.2}
    \setlength{\tabcolsep}{5pt}
\begin{center}
\begin{tabular}{l c c c c}
\hline\hline
\multirow{2}{*}{Event} & \multicolumn{2}{c}{$\log \mathcal{B}$} & \multicolumn{2}{c}{$p$value} \\
\cmidrule(lr){2-3} \cmidrule(lr){4-5}
 & New Likelihood & Old Likelihood  & New Likelihood & Old Likelihood \\
\hline
GW150914 ($114$\,s) &$-3.19$ & $-2.09$ & $0.77$ & $0.61$ \\
GW150914 ($57$\,s) &$-3.01$ & $-2.21$ & $0.59$ & $0.18$ \\
GW231226 ($145$\,s) &$-4.14$ & $-3.41$ & $1.00$ & $0.97$ \\
GW231226 ($72$\,s) &$-3.12$ & $-3.07$ & $0.20$ & $0.28$ \\
GW250114 ($58$\,s) &$-3.68$ & $-3.20$ & $0.93$ & $0.35$ \\

\hline\hline

\end{tabular}
\caption{Observed log Bayes factor and corresponding $p$value for the postmerger search of the three events.
}
\label{tab:p-value}
\end{center}
\end{table}

For the postmerger search,
we first select a data segment of duration $T$ that begins at a specified time after the peak strain of the main event (see Table~\ref{tab:parameterevent} for the exact start times). The start time is chosen so that the \ac{BH} \acp{QNM} in the early ringdown have damped sufficiently, while still being early enough to capture the dominant contribution of the long-lived \acp{QNM} of the \ac{UCO}. We then perform
a hybrid frequentist-Bayesian analysis by treating the log Bayes factor as a frequentist detection statistic. The significance of the observed Bayes factor is then  assessed via its $p$value with respect to the background distributions, i.e.
\begin{equation}
    p\text{-value} = \frac{N(\text{background} \ge \text{event})}{N(\text{total})}
\end{equation}
Table~\ref{tab:p-value} lists the observed log Bayes factor and the corresponding $p$values for the three events.
The $p$value for the shorter-duration GW150914 case analyzed with the old likelihood is consistent with our earlier search results, even though the present search explores a broader parameter space. The smallest $p$value among all configurations, $\sim18\%$, also occurs for this case. Overall, the search results for all three events are consistent with background distributions, and there is no clear evidence of the long-lived \acp{QNM} in the data.

\begin{figure}[htbp]
	\centering
	\includegraphics[width=16cm]{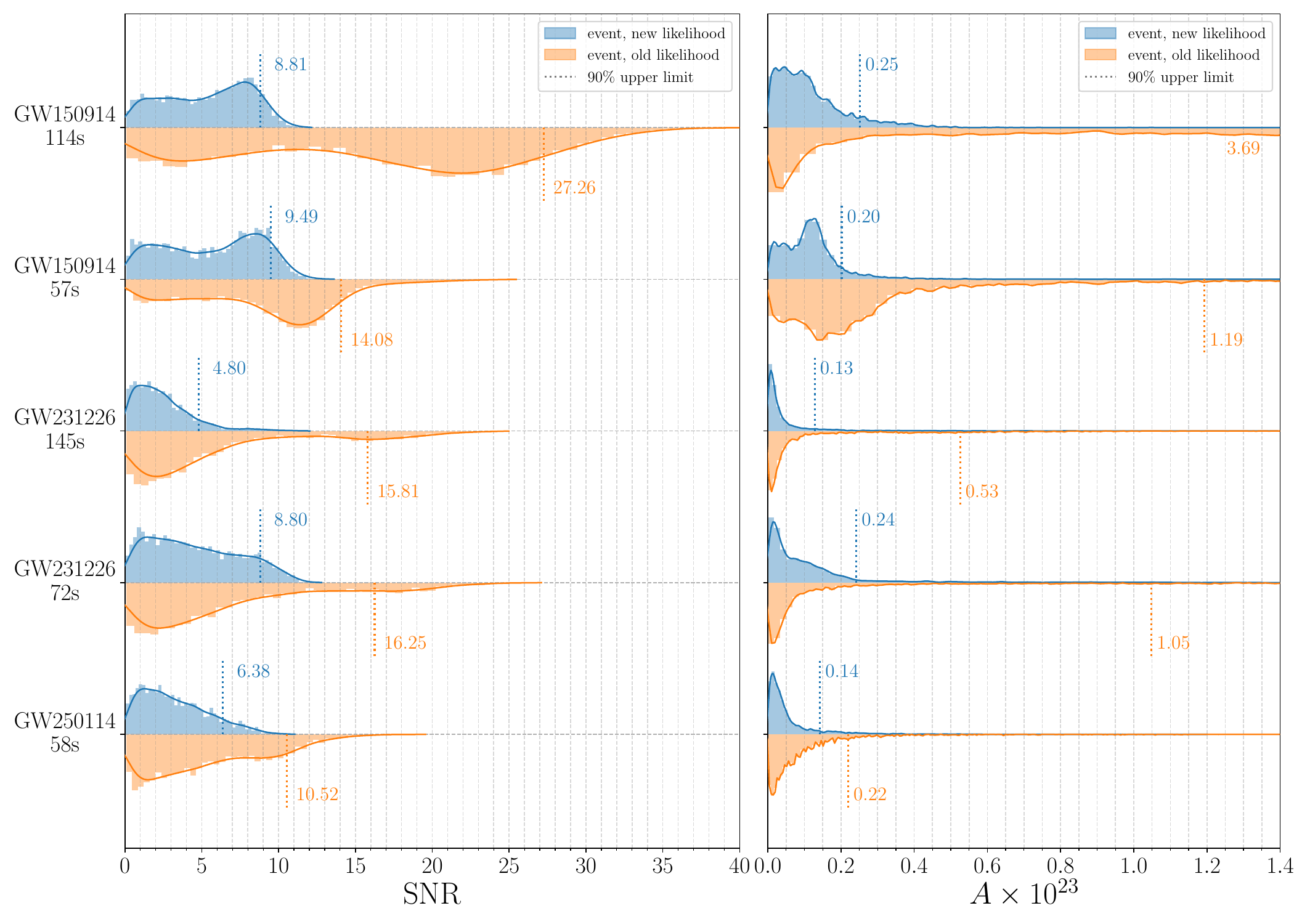}
	\caption{\label{fig:violinevent}{Postmerger search results for the three events. Left: the posterior distributions of the inferred network \ac{SNR}. Right: the posterior distributions of the average initial time-domain amplitude $A=A^\prime/\tau$. Blue and orange denote the new and old likelihood, respectively; vertical dotted lines mark the 90\% upper limits.}}
\end{figure}

With no evidence for the target signal, we constrain its strength by setting upper limits on the network \ac{SNR} and amplitude of the \ac{UniEw} search template, which approximates the \acp{QNM} at leading order.  The left panel of Figure~\ref{fig:violinevent} shows the posteriors and upper limits of the inferred \ac{SNR} from the postmerger search of the three events. The new likelihood gives more robust and typically stronger limits than the old one when the duration is varied. Among all searches, GW231226 ($T=145$\,s) and GW250114 ($T=58$\,s) give the two strongest constraints, with $\mathrm{SNR}_{90\%}\approx4.8$ and $6.4$, respectively; their comparatively low log Bayes factors---possibly reflecting a downward fluctuation in the postmerger segment---help explain this result.
The same analysis also places tight bounds on the average initial time-domain strain amplitude $A$ of the targeted long-lived QNMs,
yielding event-specific $A_{90\%}\approx1.3\times 10^{-24}$ for GW231226 and $1.4\times 10^{-24}$ for GW250114, as shown in the right panel of Figure~\ref{fig:violinevent}.
Note that the main-event strain amplitudes differ between GW231226 and GW250114. Therefore, the similar $A_{90\%}$ values obtained for the two events imply different constraining power on the physical properties of echo-generating sources. The high ringdown SNR of GW250114 makes it particularly informative for extracting bounds on source properties from the derived limits.
These results therefore place model-agnostic limits on the characteristic long-lived \acp{QNM} that govern late-time echoes in \ac{LVK} data,
complementing existing searches for other echo signatures. Nonetheless, interpreting these bounds in terms of the physical properties of echo-generating sources will require additional model-dependent assumptions and is left for future work.

\section{Summary}\label{sec:summary}

Gravitational wave echoes provide a unique probe of the near-horizon structure of \acp{UCO}, extending beyond standard \ac{BH} spectroscopy. To circumvent the theoretical uncertainties surrounding specific waveform morphologies, model-agnostic search methods are essential for robustly identifying potential echo signals in gravitational wave data.
In this work, we systematically improve the real-data search pipeline---targeting the long-lived \acp{QNM} of late-time echoes---based on our earlier studies~\cite{Ren:2021xbe,Wu:2023wfv}. Key upgrades include a generalized phase-marginalized likelihood that coherently combines all data for each \ac{QNM} across the detector network, an optimized iterative notching procedure, and refined parameter settings for Bayesian inference.
We validated this pipeline on the O1 background data around GW150914, which represent the most challenging noise conditions.
The notching procedure effectively suppresses instrumental lines. Furthermore, extensive injection tests with the previously used benchmark echo waveform confirm that the search performance follows the same trend as in Gaussian noise, and that signal parameters can be reliably recovered from real data. These results demonstrate the robustness of the pipeline in searching for the target long-lived \acp{QNM} associated with echoes under realistic observing conditions.

Applying this pipeline to three \ac{CBC} events with high ringdown \ac{SNR}---GW150914 (O1), GW231226 (O4a), and the recently reported O4 event GW250114---we performed a comprehensive search for postmerger echoes. We first examined the background noise properties around each event. While the distributions of the log Bayes factor are largely consistent with stationary Gaussian noise after notching, a tail of noise outliers with high values remains. This reveals the presence of residual non-Gaussian features beyond the obvious instrumental lines, including nonpersistent narrow-line structures and, in one case, a broad, quasiperiodic, slowly varying artifact whose origin remains unclear (see Appendix~A for details).

For the postmerger search itself, our analysis found no statistically significant evidence for gravitational wave echoes in any of the three events, with the observed log Bayes factors all consistent with background noise distributions.
Consequently, we placed upper limits on the strength of the long-lived \acp{QNM} from the inferred posteriors. Notably, the two O4 events yield the tightest 90\% upper limits obtained so far, constraining the network \ac{SNR} to $\sim5$ and the average
initial time-domain strain amplitude of the long-lived \acp{QNM} to $\sim 10^{-24}$.
These robust, model-agnostic constraints on the characteristic \acp{QNM} are the result of both the improved search method and the higher quality O4 data.

The increasing sensitivity and number of detected binary \ac{BH} mergers in future observing runs will enable deeper investigations into the nature of compact objects. To address the computational challenges of analyzing larger datasets and wider parameter ranges, future work could explore machine-learning techniques, such as normalizing flows~\cite{Dax:2021tsq,Dax:2022pxd,Langendorff:2022fzq,Polanska:2024zpn}, to accelerate the Bayesian inference process. This advancement would facilitate rapid and efficient searches for gravitational wave echoes across extensive datasets, enhancing our ability to test fundamental physics in the strong-field regime.

This research has made use of data or software obtained from the Gravitational Wave Open Science Center (gwosc.org), a service of the LIGO Scientific Collaboration, the Virgo Collaboration, and KAGRA.

\begin{acknowledgments}
We would like to thank Pengyuan Gao for communications on the echo waveform modeling and Huai-Ke Guo for discussions regarding LVK data quality. Q.G.H. is supported by the grants from National Natural Science Foundation of China (Grant No.~12547110, 12475065, 12447101) and the China Manned Space Program with grant no. CMS-CSST-2025-A01. J. R. is supported in part by the National Natural Science Foundation of China (Grant No.~12275276).
This material is based upon work supported by NSF's LIGO Laboratory which is a major facility fully funded by the National Science Foundation, as well as the Science and Technology Facilities Council (STFC) of the United Kingdom, the Max-Planck-Society (MPS), and the State of Niedersachsen/Germany for support of the construction of Advanced LIGO and construction and operation of the GEO600 detector. Additional support for Advanced LIGO was provided by the Australian Research Council. Virgo is funded, through the European Gravitational Observatory (EGO), by the French Centre National de Recherche Scientifique (CNRS), the Italian Istituto Nazionale di Fisica Nucleare (INFN) and the Dutch Nikhef, with contributions by institutions from Belgium, Germany, Greece, Hungary, Ireland, Japan, Monaco, Poland, Portugal, and Spain. KAGRA is supported by the Ministry of Education, Culture, Sports, Science and Technology (MEXT), Japan Society for the Promotion of Science (JSPS) in Japan; National Research Foundation (NRF) and Ministry of Science and ICT (MSIT) in Korea; and Academia Sinica (AS) and National Science and Technology Council (NSTC) in Taiwan.

\end{acknowledgments}

\section*{Data availability}\label{sec:data}
The strain data used in this analysis are openly available in Refs. \cite{LIGOScientific:2019lzm,KAGRA:2023pio,LIGOScientific:2025snk}. The analysis pipeline, figure reproduction scripts, and summary posterior data are available in the GitHub repository~\cite{github:echomase}.

\appendix

\section{Non-Gaussian artifacts in the detector noise}
\label{app:line}

The main non-Gaussian artifacts that affect our search for long-lived \acp{QNM} associated with echoes are narrow instrumental lines. Public observing-run data releases typically include
a catalog of such lines, identifying spectral features that are confidently of instrumental origin and can therefore be safely removed in astrophysical searches. A detailed study of these lines and their sources is given in Ref.~\cite{LSC:2018vzm}. In our analysis, we notch only a small subset of the prominent cataloged lines, making the list quite short. The most prominent lines are well understood; they include power-mains harmonics (at multiples of 60\,Hz), test-mass and beam-splitter violin modes, and others.

\begin{figure}[!htbp]
    \centering
     \includegraphics[width=5.8cm]{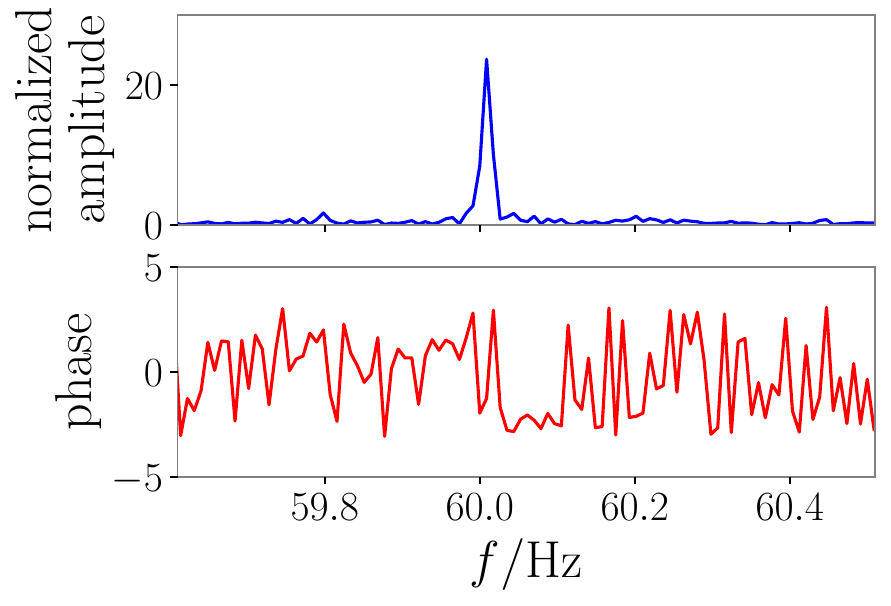}\;
     \includegraphics[width=5.8cm]{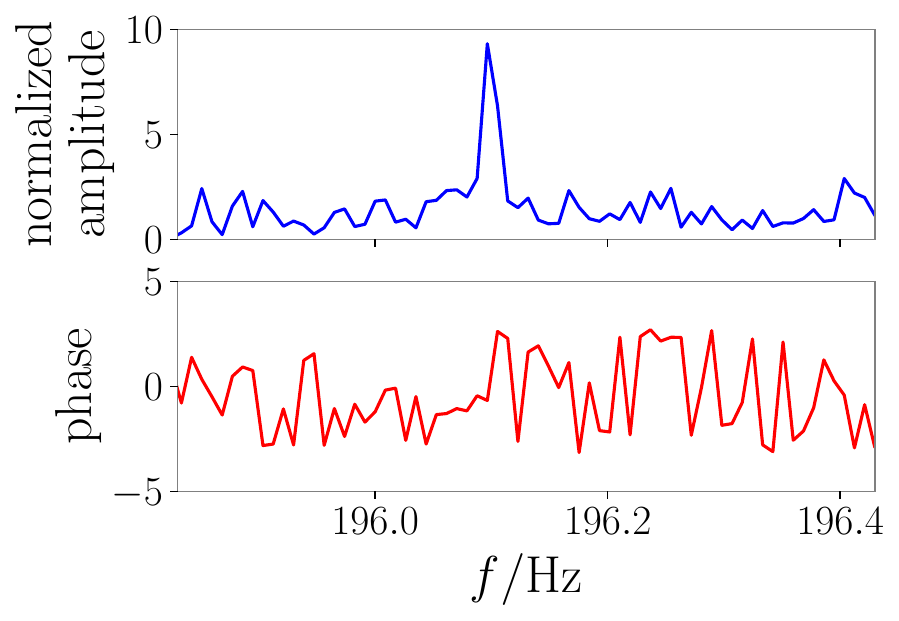}\;
     \includegraphics[width=5.8cm]{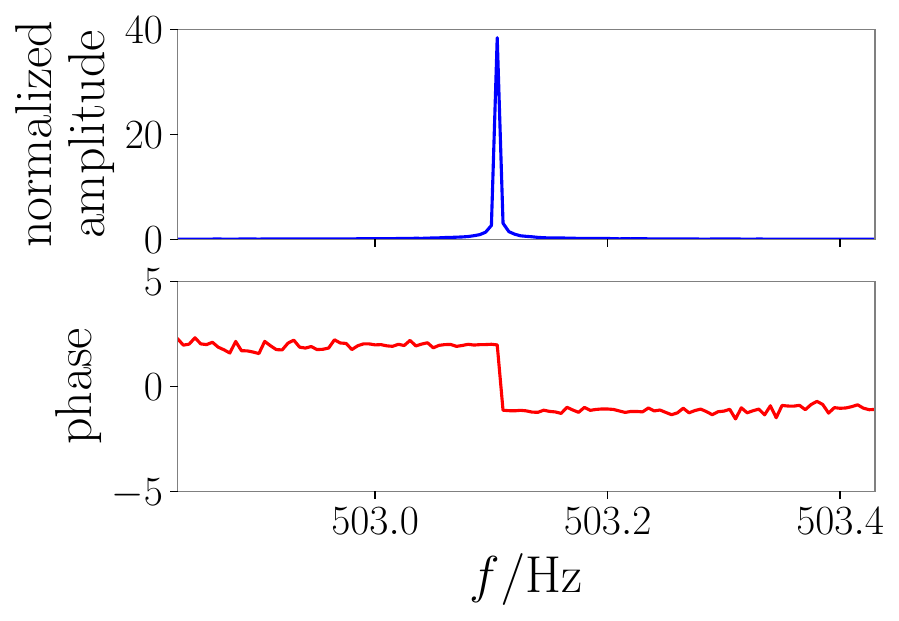}
    \caption{Normalized amplitude and relative phase for three representative examples of prominent instrumental lines on the Livingston detector of LIGO with $T=114\,$s. Left: power mains harmonics; middle: OMC length dither; right: test-mass violin mode.}
    \label{fig:both}
\end{figure}

Most instrumental lines are characterized by a narrow spectral peak, which can overlap substantially with the Lorentzian profile of the \acp{QNM}, whereas their phase evolution varies more with their physical origin. Figure~\ref{fig:both} shows the amplitude and relative phase for several prominent lines. All amplitudes are not far from the Lorentzian shape given in Eq.~(\ref{eq:abshf}), and thus are captured by the old likelihood. In phase, only the high-frequency test-mass violin mode closely follows the signal model of Eq.~(\ref{eq:arghf}); the other lines show no consistent phase pattern above the noise. Because our search is limited to frequencies below 300\,Hz for the three events, no lines with a similar phase feature appear in the band. This explains why the new likelihood efficiently suppresses line contamination and why the tail of noise outliers in Figure~\ref{fig:notchcompare} is much smaller for the new likelihood than for the old one before notching.

\begin{figure}[!htbp]
    \centering
\begin{subfigure}[b]{7.5cm}
        \centering
        \includegraphics[width=7.5cm]{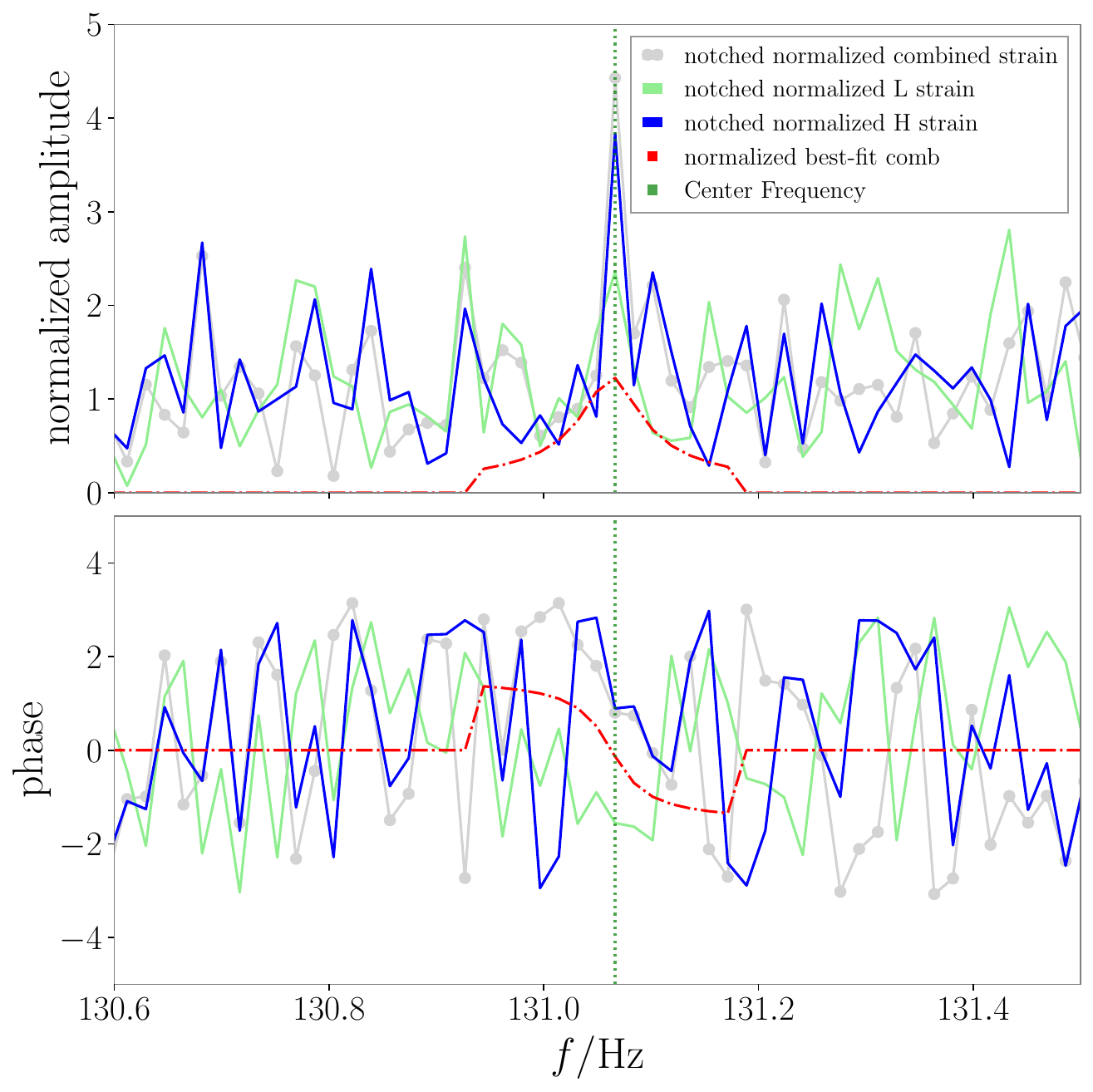}
        \caption{}
        \label{fig:line_a}
    \end{subfigure}
    \hspace{0.2cm}
    \begin{subfigure}[b]{7.5cm}
        \centering
        \includegraphics[width=7.5cm]{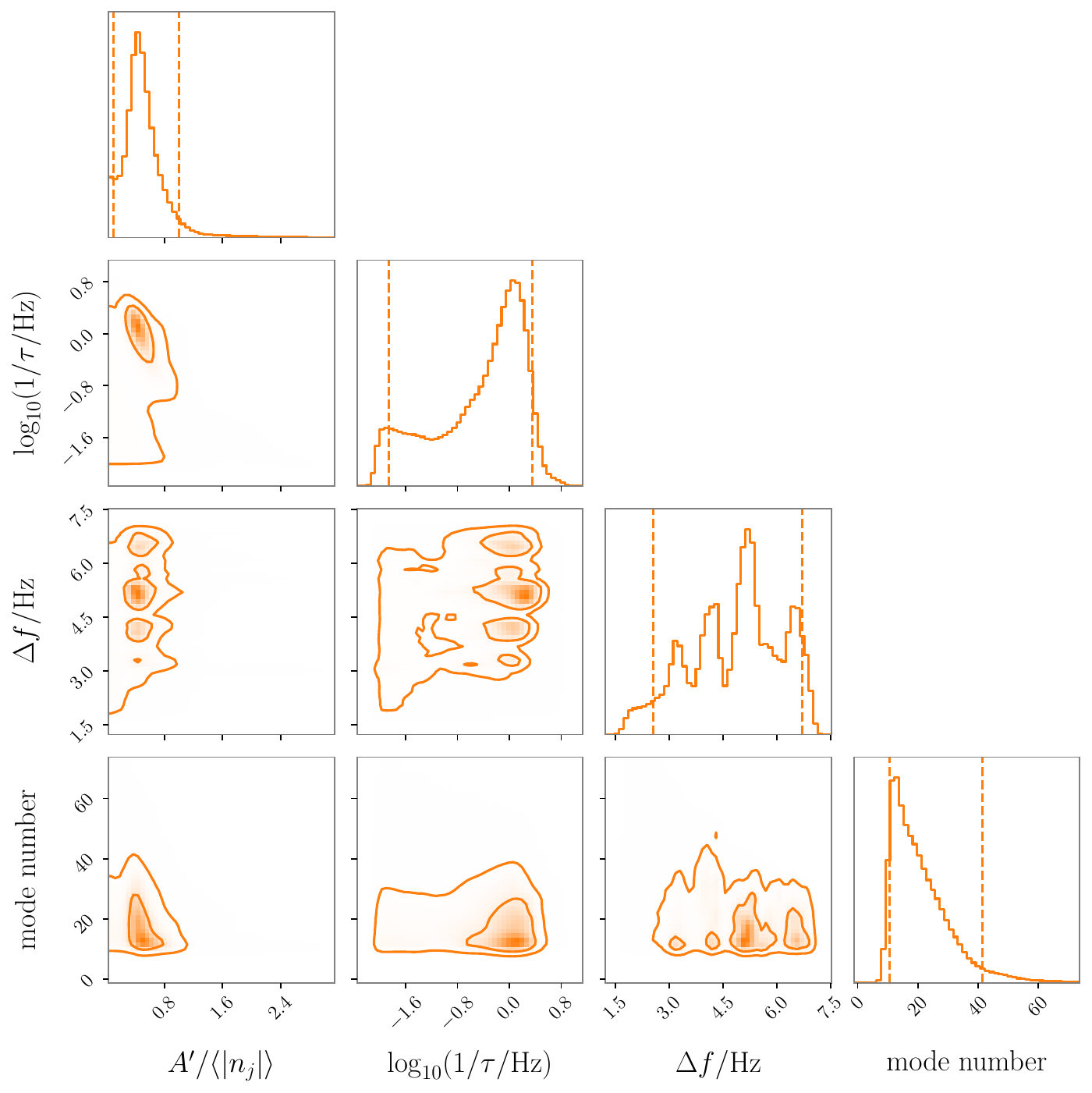}
        \caption{}
        \label{fig:line_b}
    \end{subfigure}

    \vspace{0.3cm}
    \begin{subfigure}[b]{15cm}
        \centering
        \includegraphics[width=15cm,height=5cm]{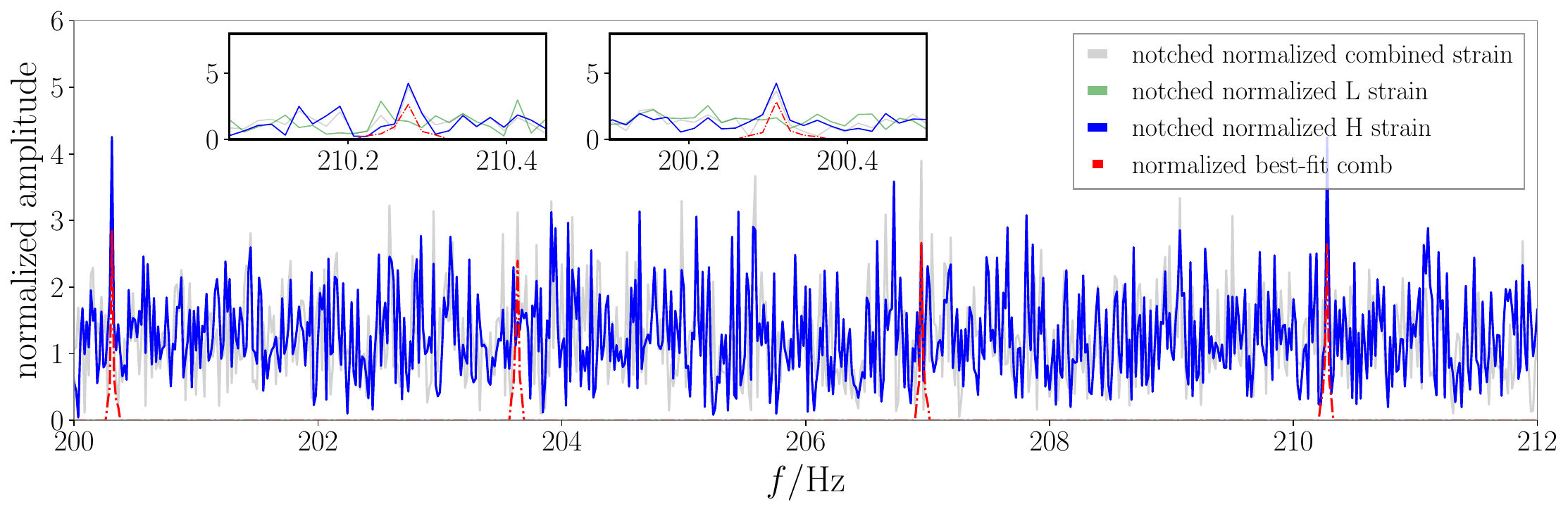}
        \caption{}
        \label{fig:line_c}
    \end{subfigure}
    \caption{Illustration of non-Gaussian artifacts identified in detector noise. (a) Example of a narrow spectral line in GW150914 (duration 57\,s) identified by the new likelihood; shown is the noise realization yielding the highest $\log \mathcal{B}$ ($[t_\mathrm{GPS}-993,t_\mathrm{GPS}-936]$\,s). (b) Corner plot of the overall posterior for GW150914 (duration 114\,s) obtained with the old likelihood ($\langle|n_j|\rangle$ is set to $6.3\times 10^{-23}/\sqrt{2}$ to account for the combined PSD $\tilde{P}_j$, making it directly comparable with the single-detector case). (c) Example of narrow spectral lines identified by the old likelihood in GW250114 (duration 58\,s) for the highest $\log \mathcal{B}$ case ($[t_\mathrm{GPS}+710,t_\mathrm{GPS}+768]$\,s). The reference GPS times $t_\mathrm{GPS}$ of GW150914 and GW250114 are shown in Table~\ref{tab:parameterevent}.}
    \label{fig:newlikelihoodline}
\end{figure}

Interestingly, after notch filtering, our search algorithm detects residual non-Gaussian artifacts in the detector noise. The background search results in Figure~\ref{fig:violinbackground} reveal noise outliers in several cases.
For the noise around GW150914, the new likelihood identifies a few outliers that correspond to narrow spectral lines in both duration settings. Figure~\ref{fig:line_a} displays an example from the noise realization that yields the highest log$\mathcal{B}$. In this case, only the Hanford detector shows a clear peak at the inferred frequency, and the phase variation near the peak does not precisely follow the signal template. However, the overall phase evolution of the combined two-detector data closely resembles the signal trend, thereby giving a large contribution to the new likelihood and producing well-behaved corner plots similar to those of an injected signal, with a best-fit \ac{SNR} around 8. This feature does not appear in other segments or other durations, demonstrating that it is not a persistent instrumental line and thus would not be listed in the line catalog. The case illustrates how coincidental phase alignment can occur and highlights the ability of our algorithm to capture such subtle, transient features.
For GW250114, the old likelihood also identifies a few noise outliers corresponding to narrow, Lorentzian-shaped spectral lines. In the case with the highest $\log \mathcal{B}$, two distinct lines are found as shown in Figure~\ref{fig:line_c}, both appearing exclusively in the Hanford detector. As with the previous examples, these are nonpersistent features. Because a true echo signal would be coherent across the detector network, such single-detector lines are inconsistent with \acp{QNM} and are thus excluded as potential detections.
Finally, for GW150914 with a longer duration, the old likelihood identifies outliers associated with a broad, quasiperiodic, and slowly varying trend that recurs in many noise realizations. The corner plot of the overall posterior in Figure~\ref{fig:line_b} reveals the detailed shape of this structure, which is clearly distinct from the narrow target lines. Particularly, its best-fit width is roughly an order of magnitude larger than in previous cases, producing a secondary peak in the \ac{SNR} posterior around 20. Such artifacts were not seen in earlier studies, likely because of the different segment duration and the enlarged prior on the width.  The origin of this wide structure requires further investigation.

\section{Search results for pure Gaussian noise } \label{app:gaussian}
In an ideal scenario, the noise of gravitational-wave detectors would be stationary and Gaussian. In practice, however, non-Gaussian effects are common. For our search targeting long-lived \acp{QNM}, the dominant contaminants are instrumental or environmental lines (detailed in Appendix~\ref{app:line}). To further understand these non-Gaussian artifacts, we revisit the background search results for pure Gaussian noise presented in our previous work~\cite{Wu:2023wfv}, which serve as a clean reference for comparison with background results obtained from actual observational data.

\begin{figure}[htbp]
    \centering
\includegraphics[width=8.5cm]{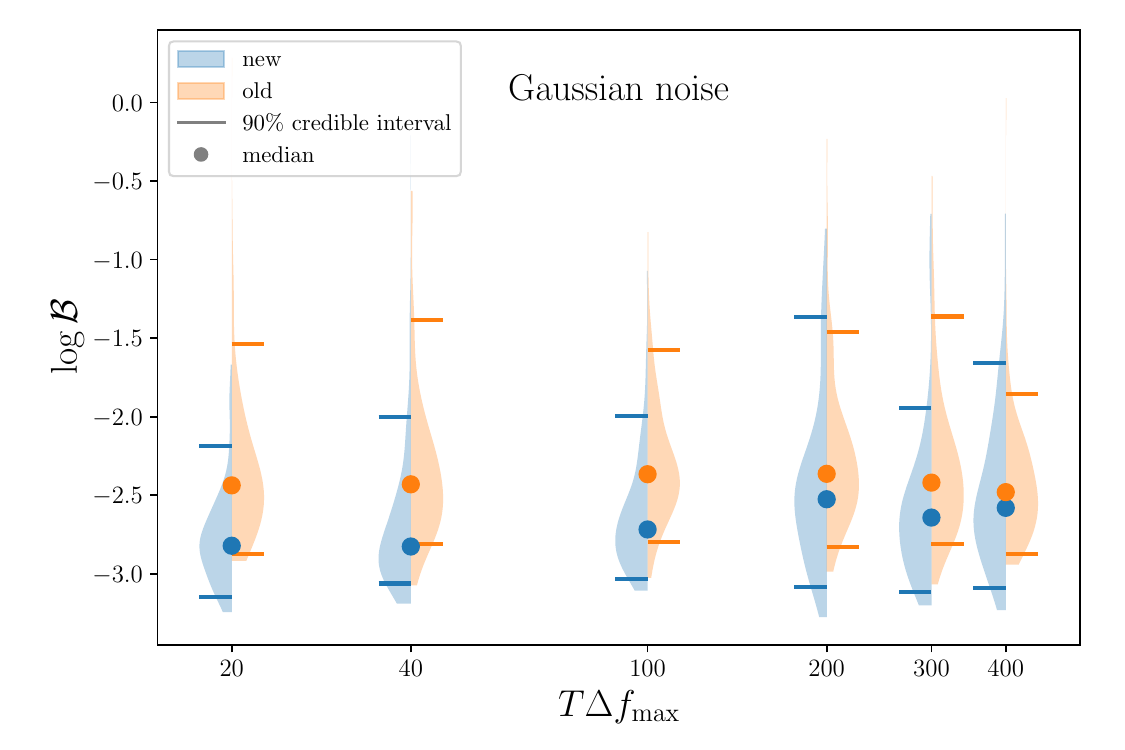}
\includegraphics[width=8.5cm]{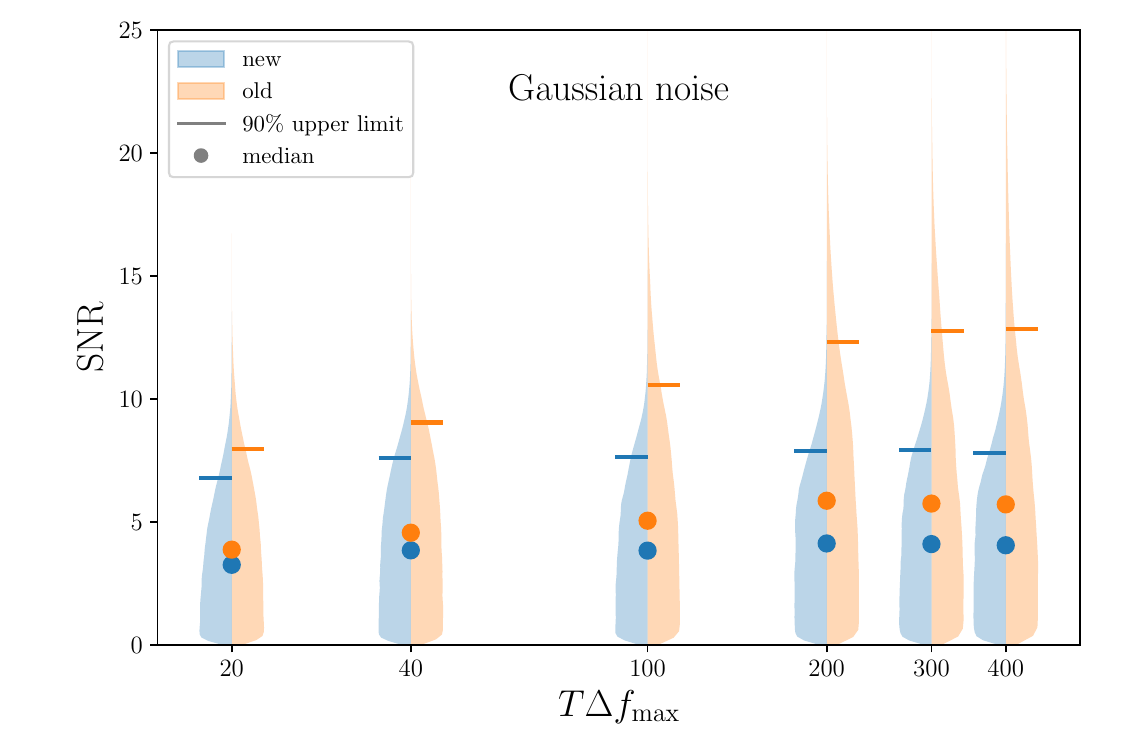}
    \caption{Background log Bayes factor distributions (left panel) and the overall posterior distributions of the inferred network \ac{SNR} (right panel) of Gaussian noise, as a function of the time-frequency product $T\Delta f_{\rm max}$. }
    \label{fig:gussian_SNR_logb}
\end{figure}

\begin{figure}[htbp]
    \centering
\includegraphics[width=16cm]{SNR_gaussian_cases.pdf}

    \caption{The inferred network \ac{SNR} posteriors for Gaussian noise using two likelihoods in the case of $T\Delta f_{\rm max}=400$. The top and bottom panels correspond to the new and old likelihoods, respectively. Each panel displays the posterior for the noise realization with the maximum $\log \mathcal{B}$ (red), the minimum $\log \mathcal{B}$ (green), and the overall posterior combining all realizations (blue for new likelihood and orange for old likelihood). }
    \label{fig:single GN event SNR}
\end{figure}

The background search results for Gaussian noise are shown in Figure~\ref{fig:gussian_SNR_logb}. The left panel for the log Bayes factor is taken from the bottom panel of Figure~18 in Ref.~\cite{Wu:2023wfv}. Results obtained with the new likelihood are slightly more negative than those with the old likelihood, but overall both remain relatively stable as the duration varies. The right panel, showing the inferred \ac{SNR} distribution for Gaussian noise, is new. It consistently peaks around zero and then gradually decreases. As the time duration increases, the tail of the \ac{SNR} distribution extends further, particularly for the old likelihood. This occurs because higher frequency resolution increases the number of frequency bins in the given band, raising the chance that random noise fluctuations will statistically mimic linelike signal features.  The length of this tail can be characterized by the 90\% upper limit of the posterior. Notably, the limit derived from the new likelihood stays quite stable, whereas the limit from the old likelihood increases considerably with longer time durations. This demonstrates the advantage of the new likelihood in providing a robust upper limit on the \ac{QNM} \ac{SNR} in the absence of a signal.

Figure~\ref{fig:single GN event SNR} shows the \ac{SNR} posterior for individual noise realizations that give the maximum and minimum $\log \mathcal{B}$ values, compared with the overall posterior. Interestingly, even in Gaussian noise, the realization with the largest $\log \mathcal{B}$ produces a posterior that is noticeably away from zero and peaks at a significantly high \ac{SNR}. In contrast, the minimal $\log \mathcal{B}$ realization yields a distribution peaked closer to zero. This illustrates the intrinsic spread of the \ac{SNR} distribution and the associated uncertainty in upper limits derived from individual noise draws. Comparing the two likelihoods, the new likelihood is less affected by this noise-realization variability. In real observational data, a larger fraction of noise realizations can exhibit a distinct peak away from zero at high \ac{SNR} owing to non-Gaussian artifacts, thereby contributing to a bimodal structure in the overall \ac{SNR} distribution in Figure~\ref{fig:violinbackground}.

\begin{figure}[H]
    \centering
\includegraphics[width=8cm]{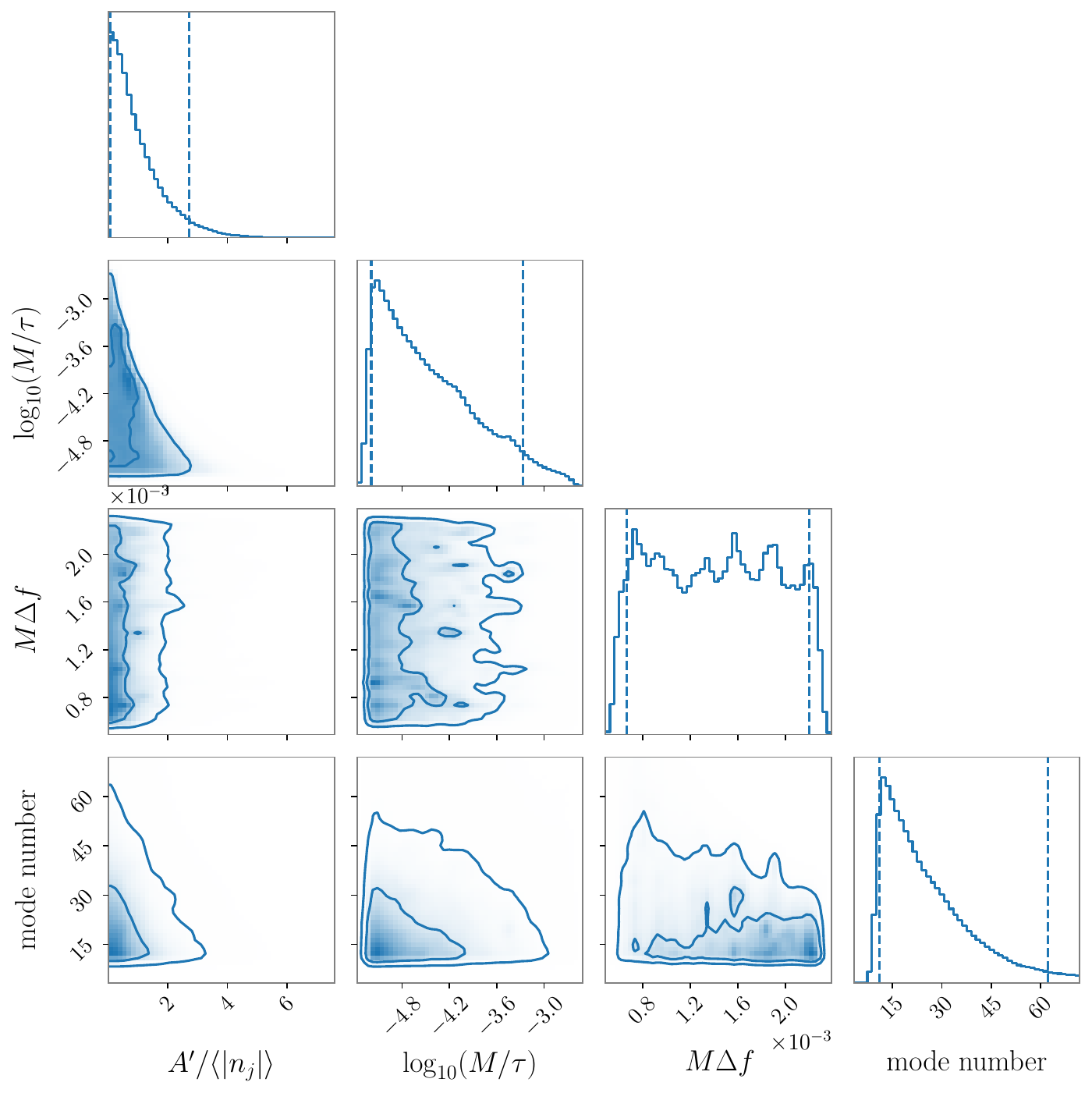}
\includegraphics[width=8cm]{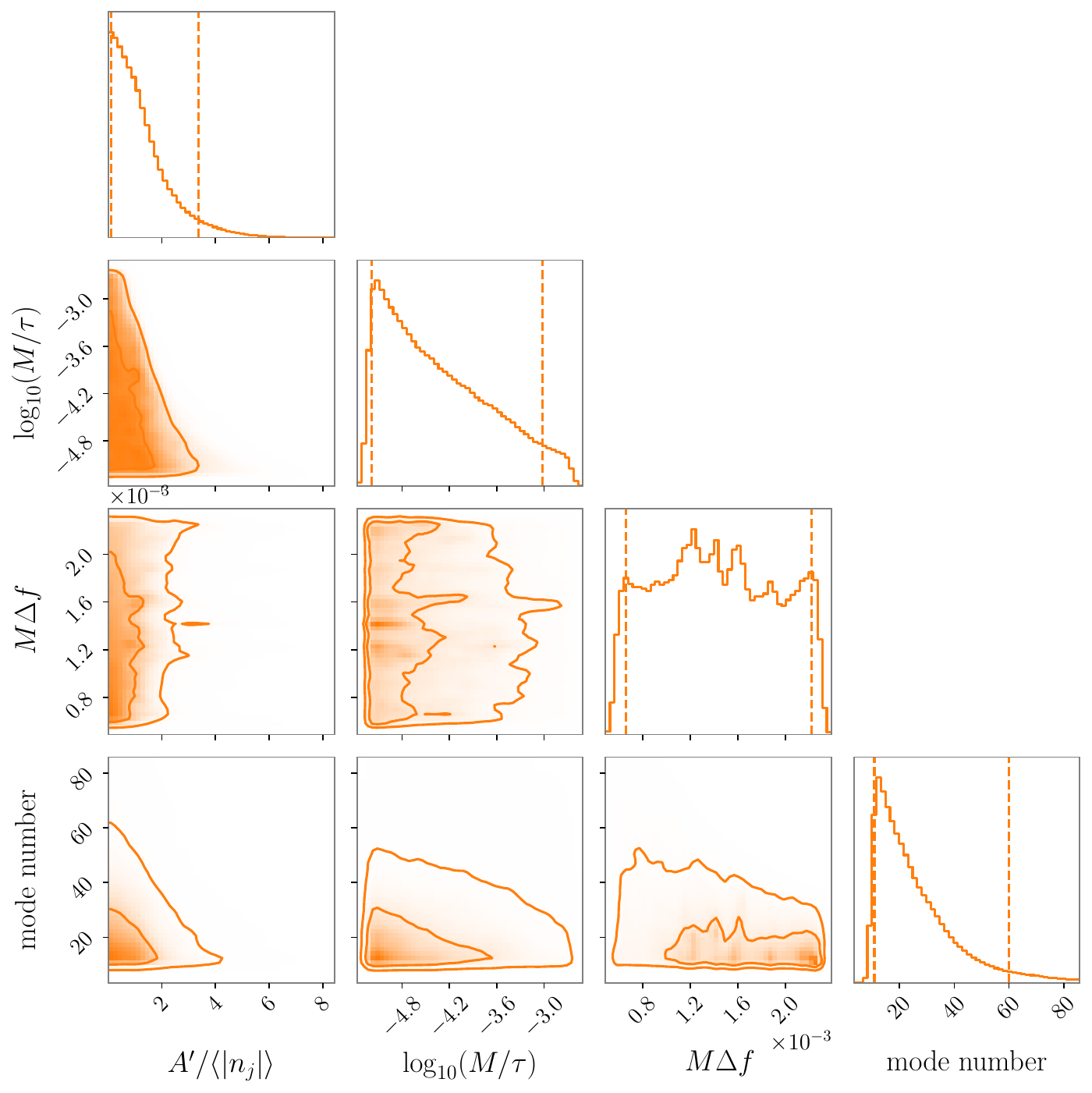}
    \caption{Corner plots of the overall posterior for the amplitude, width, spacing, and mode number derived from Gaussian noise in the case of $T\Delta f_{\rm max}=400$, using the new likelihood (left) and the old likelihood (right).
}
    \label{fig:corner}
\end{figure}

Figure~\ref{fig:corner} displays the corner plot of the overall posterior for several search parameters from Gaussian noise. As expected, the amplitude and width peak near zero, while the spacing---the typically most sensitive parameter---exhibits no distinct structure. The lower error bar of the mode number is relatively small because the frequency band must contain at least ten modes, making it tightly constrained from below. This corner plot serves as a reference for comparing with real-data search results in the presence of non-Gaussian artifacts, such as those illustrated in Figure~\ref{fig:newlikelihoodline}.

\bibliography{ref}
\end{document}